\documentclass[11pt,letter]{article}

\usepackage[margin=1in]{geometry}

\usepackage{amsmath}
\usepackage{amssymb}
\usepackage{mathtools}

\usepackage{booktabs}
\usepackage{longtable}
\usepackage{array}
\usepackage{multirow}
\usepackage{float}
\usepackage{tabularx}
\usepackage{adjustbox}
\usepackage{makecell}

\usepackage[table]{xcolor}

\AtBeginEnvironment{longtable}{\footnotesize}

\AtBeginEnvironment{longtable}{\hyphenpenalty=10000\exhyphenpenalty=10000}

\newcolumntype{L}[1]{>{\raggedright\arraybackslash}p{#1}}
\newcolumntype{C}[1]{>{\centering\arraybackslash}p{#1}}
\newcolumntype{R}[1]{>{\raggedleft\arraybackslash}p{#1}}

\usepackage{graphicx}

\usepackage{pdflscape}

\usepackage[colorlinks=true,
            linkcolor=blue,
            urlcolor=blue,
            citecolor=blue,
            pdfborder={0 0 0}]{hyperref}

\usepackage{color}
\usepackage{fancyvrb}

\DefineVerbatimEnvironment{Highlighting}{Verbatim}{commandchars=\\\{\}}
\usepackage{framed}
\definecolor{shadecolor}{RGB}{248,248,248}
\newenvironment{Shaded}{\begin{snugshade}}{\end{snugshade}}

\newcommand{\CommentTok}[1]{\textcolor[rgb]{0.56,0.35,0.01}{\textit{#1}}}

\newcommand{\ControlFlowTok}[1]{\textcolor[rgb]{0.13,0.29,0.53}{\textbf{#1}}}

\newcommand{\DecValTok}[1]{\textcolor[rgb]{0.00,0.00,0.81}{#1}}

\newcommand{\FloatTok}[1]{\textcolor[rgb]{0.00,0.00,0.81}{#1}}

\newcommand{\KeywordTok}[1]{\textcolor[rgb]{0.13,0.29,0.53}{\textbf{#1}}}
\newcommand{\NormalTok}[1]{#1}
\newcommand{\OperatorTok}[1]{\textcolor[rgb]{0.81,0.36,0.00}{\textbf{#1}}}

\newcommand{\StringTok}[1]{\textcolor[rgb]{0.31,0.60,0.02}{#1}}
\newcommand{\VariableTok}[1]{\textcolor[rgb]{0.00,0.00,0.00}{#1}}

\usepackage{setspace}
\usepackage{enumitem}
\usepackage{caption}
\usepackage{needspace}  

\usepackage{titlesec}
\titleformat{\section}{\Large\bfseries}{}{0em}{}
\titleformat{\subsection}{\large\bfseries}{}{0em}{}
\titleformat{\subsubsection}{\normalsize\bfseries}{}{0em}{}

\usepackage{fvextra}
\DefineVerbatimEnvironment{Highlighting}{Verbatim}{
  breaklines,
  commandchars=\\\{\},
  samepage=true
}

\usepackage{etoolbox}
\BeforeBeginEnvironment{Shaded}{\begin{samepage}}
\AfterEndEnvironment{Shaded}{\end{samepage}}

\title{Generative AI for Stock Selection}

\author{Keywan Christian Rasekhschaffe\\{\small CodeWilling Inc.}}

\date{January 2026}

\begin{document}

\maketitle

\begin{abstract}
\noindent We study whether generative AI can automate feature discovery
in U.S. equities. Using large language models with retrieval-augmented
generation and structured/programmatic prompting, we synthesize
economically motivated features from analyst, options, and price-volume
data. These features are then used as inputs to a tabular
machine-learning model to forecast short-horizon returns. Across
multiple datasets, AI-generated features are consistently competitive
with baselines, with Sharpe improvements ranging from 14\% to 91\%
depending on dataset and configuration. Retrieval quality is pivotal:
better knowledge bases materially improve outcomes. The AI-generated
signals are weakly correlated with traditional features, supporting
combination. Overall, generative AI can meaningfully augment feature
discovery when retrieval quality is controlled, producing interpretable
signals while reducing manual engineering effort.
\end{abstract}
\newpage

\subsection{1. Introduction}\label{introduction}

Using large language models enhanced with retrieval-augmented generation
(RAG) and dynamic structured prompting (DSPy), we synthesize
economically motivated features that, when used in a gradient-boosted
tabular pipeline, deliver implementable (signals lagged one extra day)
Sharpe ratios of 1.14--1.63 in ensemble configurations across datasets,
with higher values under standard T→T+1 evaluation (up to 2.624). A
central observation concerns the role of domain knowledge accuracy: when
provided with incorrect documentation, the same AI system produces
features with negative Sharpe ratios (-0.109), while correcting the
knowledge base is associated with profitable signals (Sharpe 1.27),
indicating that retrieval quality can materially affect outcomes
(p\textless0.001).

We evaluate three distinct configurations throughout: (i) Baseline
(traditional features only), (ii) Structured Prompting (AI-generated
features from hand-engineered prompts, with RAG retrieval), and (iii)
DSPy Programmatic Prompting (AI-generated features via automated prompt
optimization, with RAG). We also report Combined models that merge
baseline and AI-generated features. This taxonomy separates the
contribution of traditional features, prompt design, and programmatic
prompting, and it allows us to quantify how retrieval quality affects
each.

The feature engineering bottleneck has long constrained systematic
equity strategies. While gradient boosting models and neural networks
have proven effective at combining existing features, the space of
potential transformations remains largely unexplored due to manual
constraints. Recent applications of large language models to finance
have focused primarily on processing unstructured text: analyzing
earnings calls, parsing news sentiment, and extracting information from
SEC filings. We take a different approach: using LLMs to generate
executable code that transforms structured financial data into trading
signals. This requires the model to understand both the economic
relationships between variables and the technical constraints of
backtesting, including proper handling of point-in-time data and
avoidance of look-ahead bias.

Our framework operates on three complementary data sources spanning a
decade of market data from 2015-2024, encompassing multiple market
regimes including the COVID volatility shock, the 2022 rate hiking
cycle, and subsequent recovery periods. The TrueBeats dataset provides
predictions of earnings surprises derived from sell-side analyst
forecasts, earnings management metrics, and time-series trends. The
SpiderRock options dataset captures derivatives market information
including implied volatility surfaces, flow estimators, and risk-neutral
metrics. The price-volume universe covers standard market data for over
2,500 securities. The framework generated over 500 unique features
across these datasets through iterative discovery processes, with each
dataset receiving several hundred candidate features that underwent
rigorous evaluation for statistical significance and economic validity.
When combined in machine learning models, these AI-generated features
produce information coefficients in the 0.003-0.010 range. The generated
features exhibit low correlations with traditional factors (0.066 to
0.147), suggesting they capture patterns that are not redundant with the
baselines.

We highlight three observations with practical relevance. First,
retrieval quality acts as a critical control variable: differences
between accurate and corrupted documentation are associated with large
changes in Sharpe (p\textless0.001). Second, requiring brief economic
rationales during feature generation is associated with higher
interpretability and modestly higher information coefficients (observed
+15-20\% vs pure code generation in our runs). Third, the DSPy
programmatic optimization approach, which automatically learns effective
prompting strategies from successful examples, showed mixed results
compared to hand-crafted structured prompting. On the SpiderRock options
dataset, DSPy achieved competitive performance under one-day
execution-lag evaluation (Sharpe 1.461), with even stronger results on
standard returns (Sharpe 2.624, see Appendix), while results varied
across other datasets, suggesting that the optimal prompting approach
may be dataset-specific. This finding highlights that automated prompt
optimization complements rather than universally replaces careful prompt
engineering. All headline results use one-day execution-lag evaluation
(F1; returns T+1 to T+2). For completeness, we also report standard
next-day returns (T-\textgreater T+1) in the appendix. We also analyze
the patterns discovered by the generation system later in the paper,
revealing systematic approaches that align with established principles
in quantitative finance.

\subsection{2. Related Work}\label{related-work}

The empirical asset-pricing literature documents that firm
characteristics explain cross-sectional variation in expected returns
(Fama and French 1993), with the set of proposed predictors expanding to
the hundreds (Green, Hand, and Zhang 2017). More recently,
machine-learning methods have improved cross-sectional forecasting by
capturing nonlinearities and interactions among characteristics that
linear models miss (Gu, Kelly, and Xiu 2020). In practice, forecast
combinations often perform robustly relative to single-model
alternatives (Rasekhschaffe and Jones 2019), motivating evaluations that
compare baseline features, newly proposed signals, and their
combinations. Taken together, this literature provides both the
motivation and the baseline for assessing whether additional signals
(here, those proposed by large language models) can add incremental
information under realistic constraints.

Parallel work examines large language models (LLMs) in financial
applications, primarily in text domains such as earnings-call analysis
and news-based sentiment (e.g., Lopez-Lira and Tang 2023; Chen et
al.~2023; Wu et al.~2023). These studies demonstrate that
general-purpose LLMs can extract economically relevant information from
unstructured sources, and finance-tuned systems (e.g., BloombergGPT; Wu
et al.~2023) illustrate domain adaptation. Our focus differs: we study
structured-data feature discovery, in which an LLM proposes explicit
transformations over tabular inputs that can be implemented directly in
quantitative pipelines. For practitioners, this distinction matters
operationally (structured features are deployable without maintaining
NLP pipelines and document stores) and methodologically, because
proposed transformations must reference available columns, respect
schema constraints, and avoid look-ahead.

Two methodological strands are particularly relevant to structured-data
discovery. First, retrieval-augmented generation (RAG) conditions the
model on external documentation during prompting (Lewis et al.~2020).
Retrieval accuracy and relevance can materially affect output quality in
tasks requiring domain expertise, with domain-specific evidence pointing
to significant sensitivity when context is inaccurate or incomplete
(e.g., Chen et al.~2023). In our setting, retrieval quality is a
first-order control variable: schema accuracy and clear field
definitions help constrain the hypothesis space and reduce invalid
proposals. Second, programmatic prompting frameworks, of which DSPy
(Khattab et al.~2023) is a representative example, formalize prompt
composition by learning from examples and feedback, reducing ad hoc
iteration. Rather than relying on static templates, these approaches
compile prompting strategies into reusable programs that emphasize
structure, constraints, and objective-aligned feedback signals.

Outside finance, recent work reports that schema-constrained prompting
can yield useful tabular transformations when models are guided by
examples, scorecards, or validation signals (e.g., Li et al.~2024; Wang
et al.~2025; Snowflake AI Labs 2024). Reported gains are typically
modest but consistent, often on the order of 10-15\% improvements in
downstream accuracy, suggesting that LLMs can act as hypothesis
generators when properly constrained and validated. Translating these
ideas to financial data requires additional safeguards: strict
point-in-time usage, explicit anti-leakage constraints, and semantic
de-duplication to avoid near-equivalent features. Moreover, evaluation
must emphasize out-of-sample testing with appropriate inference to
separate genuine signal from sampling variation and multiple-testing
risk.

Our contribution is to position LLM-assisted feature discovery within
the cross-sectional equity literature by comparing (i) baseline models
using conventional features, (ii) structured prompting that elicits
features from hand-crafted prompts, and (iii) programmatic prompting via
a compiled prompting approach. We quantify how retrieval quality
influences outcomes and assess whether AI-generated features complement
baseline signals when combined, emphasizing deployability (auditable
code and schema compliance) and backtesting hygiene (point-in-time
inputs and leakage controls). The analysis focuses on out-of-sample
performance over a common multi-year window across datasets and uses
standard statistical comparisons to align with prior work on model
evaluation in asset management. This design isolates the incremental
value of structured-data feature discovery and clarifies when automated
prompting complements, rather than replaces, careful feature
engineering.

\begin{center}\rule{0.5\linewidth}{0.5pt}\end{center}

\subsection{3. Methodology}\label{methodology}

This section details data construction, feature generation, model
estimation, portfolio formation, and inference. The design emphasizes
point-in-time inputs, forward-chaining estimation, and out-of-sample
evaluation over a common window from January 2019 through December 2024.

\subsubsection{3.1 Data and Sample
Construction}\label{data-and-sample-construction}

Our empirical tests draw on three complementary datasets that together
span the main types of information used in equity research: price and
trading activity, analyst expectations, and options market signals. All
data are aligned at daily frequency and linked using a common CWIQ\_Code
security identifier. To ensure that results are comparable across
sources, we limit our analysis to the 2,500 largest U.S. common equities
by market capitalization at each point in time. This restriction removes
microcaps - where coverage is sparse and data quality often deteriorates
- and focuses the analysis on a liquid, investable universe. We exclude
ADRs, ETFs, closed-end funds, and preferred shares. Prices are adjusted
for corporate actions, and delisting returns are included in line with
standard practice.

Unless otherwise noted, the core of our analysis evaluates predictive
signals using forward returns measured from trading day t+1 to t+2. This
choice reflects a conservative and implementable forecasting horizon for
most quantitative strategies, ensuring that signals are evaluated in a
way that would be feasible to trade in practice. For completeness, we
also provide examples of results based on returns from t to t+1 in the
appendix.

Our starting point is daily end-of-day price and volume data from
Exchange Data International (EDI). These records include open, high,
low, and close prices, as well as trading volume in both shares and
dollars. While these variables are basic, they capture the essential ebb
and flow of market activity and form the foundation of most empirical
work in equities. From this broader dataset, we extract the filtered
universe described above. This step is not just about consistency: it
also reduces noise from illiquid securities, avoids distortions from
penny stocks, and ensures that the features we generate reflect dynamics
relevant to institutional portfolios. The resulting dataset provides a
clear view of how prices and volumes evolve over time, serving as a
benchmark against which we can measure the incremental value of new
signals.

We extend this foundation with forward-looking information from the
TrueBeats dataset, produced by ExtractAlpha. TrueBeats draws on
sell-side analyst forecasts, indicators of earnings management, and
time-series trends in corporate fundamentals to estimate the likelihood
of earnings-per-share and revenue surprises. All variables are
constructed in a strictly point-in-time manner and overlap closely with
the EDI universe. The appeal of this dataset lies in its ability to
summarize investor expectations about company performance - something
that raw price data alone cannot capture. Because many cross-sectional
strategies are built on revisions and surprises in analyst expectations,
this dataset provides a natural testing ground for whether generative
models can propose features that extend or improve upon the conventional
approaches.

Finally, we incorporate information from the SpiderRock Option Features
dataset, which supplies a rich view of derivatives markets from January
2015 onward. It contains daily measures of implied volatility surfaces,
skewness, term structure, and other risk-neutral quantities that reflect
how investors price uncertainty and tail risk. It also includes data on
order flow, trading imbalances, and positioning - variables that give
insight into the sentiment and behavior of different market
participants. Options data add a crucial dimension: they represent
forward-looking views on volatility and risk that are not directly
observable in cash equities. By including this dataset, we can assess
whether LLM-generated features are capable of identifying predictive
relationships embedded in derivatives markets - relationships that might
remain hidden when focusing on prices or fundamentals alone.

Taken together, these three datasets offer a broad and diverse view of
the information available to investors. They cover fundamentals, trading
activity, and risk expectations, and they do so across multiple market
regimes between 2015 and 2024. Because all data are mapped to the same
securities and frequency, we can evaluate models on a consistent basis
and attribute differences in predictive performance to the type of
information being used rather than to differences in coverage or timing.
The unified design - combined with the consistent t+1 to t+2 return
horizon - provides a robust foundation for testing the scope and
limitations of generative feature discovery in equity markets.

\subsubsection{3.2 Experimental Design and Sample
Splits}\label{experimental-design-and-sample-splits}

We partition the sample chronologically into a development
(``discovery'') window and an out-of-sample evaluation window. The
discovery window (January 2015-December 2018) is used to propose and
screen feature candidates and to select hyperparameters; no statistics
from this window are reported as headline results. Models are estimated
and evaluated on a walk-forward basis. The out-of-sample window (January
2019-December 2024) is used exclusively for reporting portfolio
performance and statistical tests. We focus evaluation on an
implementation-lag variant in which trades are executed one day after
signal formation (t+2 execution) unless stated otherwise.

\subsubsection{3.3 Feature Generation and
Validation}\label{feature-generation-and-validation}

We implement two generation routes that feed a common validation and
evaluation pipeline: a hand-designed, prompt template (``structured
prompting'') and a programmatic route (DSPy) that starts with a simple
signature and then automatically generates a prompting program. Both
routes query OpenAI's GPT-4.1 model via the API and encode identical
constraints: the dataset schema, point-in-time usage with non-negative
lags and correctly bounded rolling windows, and a fixed function
interface that returns a single time-aligned Series keyed by security
identifier (CWIQ\_code) and date.

Both routes are conditioned on retrieved documentation from a knowledge
base comprising schemas, field definitions, and dataset descriptions
distilled from vendor white papers and documentation. The retrieved
context constrains proposals and is held fixed during discovery and
out-of-sample evaluation, except where we explicitly contrast corrupted
versus corrected context.

In the structured-prompting route, the template enumerates the
permissible input columns and operations and requests a brief economic
rationale. The template is fixed prior to out-of-sample assessment.
Proposals are retained only if the code compiles under the schema and
passes static checks. We also control for availability and redundancy:
candidates with materially sparse cross-sectional or temporal coverage
after point-in-time construction are penalized.

In the DSPy route, we begin by compiling an optimized program using
MIPRO (Multi-Instruction Prompt Optimization; Opsahl-Ong et al.~2024) on
a training corpus. The compilation process uses a scoring function that
weighs both code quality (whether candidates compile and pass validation
checks) and in-sample predictive performance measured by IC and Sharpe
ratio. Once compiled, this optimized program generates multiple feature
candidates per request. During the discovery window, we track
portfolio-level Sharpe ratios and cross-sectional information
coefficients against next-day returns, feeding these in-sample signals
back across batches to guide the variety and quality of subsequent
proposals. All headline performance metrics are reported exclusively on
the out-of-sample test period.

All criteria (implementability, point-in-time construction, sparsity
penalties, similarity suppression, diversity, and portfolio-based
feedback) are applied identically to both routes. We report Baseline,
Structured-Prompt, and DSPy variants side-by-side, and, where
informative, a Combined model.

\subsubsection{3.4 Mathematical Framework}\label{mathematical-framework}

This section formalizes the mapping from raw data to portfolio
positions. We define notation, specify the prediction model, and
describe how forecasts translate into tradable weights.

\textbf{Notation.} Let \(i \in \{1, \ldots, N_t\}\) index securities in
the cross-section at time \(t\), where \(N_t \approx 2{,}500\) after
applying our universe filters. Denote by
\(r_{i,t+1} = \ln(P_{i,t+1}/P_{i,t})\) the log return of security \(i\)
from \(t\) to \(t+1\), and by \(\mathbf{x}_{i,t} \in \mathbb{R}^p\) the
vector of \(p\) features observed at or before time \(t\).

\textbf{Feature Sets.} We partition features into baseline and
AI-generated components:

\[\mathbf{x}_{i,t} = \begin{cases}
\mathbf{x}_{i,t}^{\text{base}} & \text{(Baseline configuration)} \\
\mathbf{x}_{i,t}^{\text{AI}} & \text{(AI-only configuration)} \\
[\mathbf{x}_{i,t}^{\text{base}}; \mathbf{x}_{i,t}^{\text{AI}}] & \text{(Combined configuration)}
\end{cases}\]

where \([\cdot; \cdot]\) denotes concatenation. Baseline features
\(\mathbf{x}^{\text{base}}\) comprise the raw vendor-supplied columns:
for TrueBeats, analyst coverage counts, consensus estimates, and
earnings surprise predictions; for SpiderRock, implied volatility
surfaces, option Greeks, and flow imbalances; for the Price Universe,
price and volume fields.

AI-generated features \(\mathbf{x}^{\text{AI}}\) are \textbf{transforms
of the baseline columns}. The LLM generates Python code that takes these
raw dataset columns as inputs and applies combinations of rolling
windows, cross-sectional operations, and nonlinear transformations. The
baseline features are the vendor-supplied columns themselves; the AI
features are derived quantities computed from those same columns. All AI
features satisfy identical point-in-time constraints.

\textbf{Prediction Model.} We model expected returns as a nonlinear
function of features:

\[\mathbb{E}[r_{i,t+1} \mid \mathbf{x}_{i,t}] = f(\mathbf{x}_{i,t}; \theta)\]

where \(f(\cdot; \theta)\) is a gradient-boosted decision tree ensemble
(LightGBM) with parameters \(\theta\). Unlike linear factor models of
the form
\(r_{i,t+1} = \alpha_i + \boldsymbol{\beta}_i' \mathbf{F}_{t+1} + \varepsilon_{i,t+1}\),
which assume returns load linearly on a small set of common factors
\(\mathbf{F}\), our specification allows for arbitrary nonlinearities
and high-dimensional interactions among features. The parameters
\(\theta\) are estimated by minimizing the mean squared prediction error
over the training window:

\[\hat{\theta} = \arg\min_{\theta} \sum_{t \in \mathcal{T}_{\text{train}}} \sum_{i=1}^{N_t} \left( r_{i,t+1} - f(\mathbf{x}_{i,t}; \theta) \right)^2 + \Omega(\theta)\]

where \(\Omega(\theta)\) is a regularization penalty controlling tree
depth and leaf weights. The model is re-estimated on a rolling basis as
described in Section 3.5.

\textbf{Cross-Sectional Standardization.} Raw predictions
\(\hat{r}_{i,t+1} = f(\mathbf{x}_{i,t}; \hat{\theta})\) are transformed
to standardized scores within each cross-section:

\[\alpha_{i,t} = \frac{\hat{r}_{i,t+1} - \bar{r}_t}{\sigma_t}, \quad \text{where } \bar{r}_t = \frac{1}{N_t}\sum_{j=1}^{N_t} \hat{r}_{j,t+1}, \quad \sigma_t = \sqrt{\frac{1}{N_t}\sum_{j=1}^{N_t}(\hat{r}_{j,t+1} - \bar{r}_t)^2}\]

Scores are then winsorized at \(\pm 3\) standard deviations:
\(\tilde{\alpha}_{i,t} = \max(-3, \min(3, \alpha_{i,t}))\).

\textbf{Portfolio Construction.} Standardized scores map to portfolio
weights via a linear transformation that enforces dollar neutrality and
unit gross leverage:

\[w_{i,t} = \frac{\tilde{\alpha}_{i,t}^+}{\sum_{j} \tilde{\alpha}_{j,t}^+} - \frac{\tilde{\alpha}_{i,t}^-}{\sum_{j} \tilde{\alpha}_{j,t}^-}\]

where \(\tilde{\alpha}^+ = \max(\tilde{\alpha}, 0)\) and
\(\tilde{\alpha}^- = \max(-\tilde{\alpha}, 0)\) denote the positive and
negative parts. This ensures \(\sum_i w_{i,t} = 0\) (dollar neutrality)
and \(\sum_i |w_{i,t}| = 2\) (unit leverage on each side). The realized
portfolio return is:

\[R_{t+1}^{\text{port}} = \sum_{i=1}^{N_t} w_{i,t} \cdot r_{i,t+1}\]

\textbf{Performance Evaluation.} We assess predictive accuracy using the
cross-sectional information coefficient:

\[\text{IC}_t = \text{Corr}_{\text{Spearman}}\left( \{\alpha_{i,t}\}_{i=1}^{N_t}, \{r_{i,t+1}\}_{i=1}^{N_t} \right)\]

and risk-adjusted performance via the annualized Sharpe ratio:

\[\text{SR} = \frac{\sqrt{252} \cdot \bar{R}}{\sigma_R}, \quad \text{where } \bar{R} = \frac{1}{T}\sum_{t=1}^{T} R_t^{\text{port}}, \quad \sigma_R = \sqrt{\frac{1}{T}\sum_{t=1}^{T}(R_t^{\text{port}} - \bar{R})^2}\]

\subsubsection{3.5 Model Estimation}\label{model-estimation}

Forecasts are produced with gradient-boosted decision trees (LightGBM).
We consider three configurations: (i) Baseline models using only vendor
or conventional features; (ii) AI-only models using the accepted
LLM-generated features; and (iii) Combined models that merge baseline
and AI-generated features. Hyperparameters are selected in the discovery
window using a rolling expanding scheme with nested cross-validation;
the selected configuration is then fixed for out-of-sample evaluation.
All input features are cross-sectionally standardized each day.

\subsubsection{3.6 Portfolio Formation}\label{portfolio-formation}

Each day, we transform model scores into a tradable long-short
portfolio. Scores are cross-sectionally z-scored and winsorized at +/-3
standard deviations. We form dollar-neutral positions with weights
proportional to standardized scores and rescale to unit gross leverage.
Portfolios are rebalanced daily. When multiple model variants are
combined, we average their standardized scores prior to forming a single
portfolio. We report results gross of transaction costs in the main text
and provide a separate analysis of net performance under realistic cost
assumptions.

\subsubsection{3.7 Performance Metrics}\label{performance-metrics}

We report daily Sharpe ratios, Spearman rank information coefficients
(IC), hit rates, and maximum drawdowns. IC summarizes cross-sectional
association between signals and next-day returns. For the
implementation-lag variant, we compute the same statistics with signals
executed one day later.

\subsubsection{3.8 Statistical Inference}\label{statistical-inference}

Standard errors for daily Sharpe ratios and IC account for serial
correlation using heteroskedasticity- and autocorrelation-consistent
(HAC) estimators (Newey-West). Where noted, we provide bootstrap
confidence intervals for selected statistics using a stationary
bootstrap with block length chosen by rule of thumb.

\subsubsection{3.9 Robustness Analyses}\label{robustness-analyses}

We examine stability across size segments by repeating the evaluation in
market-capitalization terciles and across market regimes defined by
volatility (e.g., VIX quintiles). We also evaluate the
implementation-lag variant to gauge near-term decay and provide model
interpretability summaries (feature importance) for the AI-only and
Combined configurations. Conclusions are unchanged across these segments
unless otherwise noted.

\subsubsection{3.10 Transaction Costs}\label{transaction-costs}

We evaluate net-of-cost performance using a position-level transaction
cost model that accounts for both explicit and implicit trading
frictions.

\textbf{Spread Cost.} For each position, we compute the half bid-ask
spread as a fraction of the mid price:

\[c_{\text{spread},i} = \frac{\text{ask}_i - \text{bid}_i}{2 \cdot \text{mid}_i}\]

This cost is incurred on each side of the trade (entry and exit).

\textbf{Market Impact.} We model price impact using a square-root
specification common in the market microstructure literature (Almgren
and Chriss 2000; Grinold and Kahn 2000):

\[c_{\text{impact},i} = k \cdot \sigma_i \cdot \sqrt{\frac{Q_i}{\text{ADV}_i}}\]

where \(\sigma_i\) is the 20-day realized volatility, \(Q_i\) is the
dollar position size, \(\text{ADV}_i\) is the 21-day median dollar
volume, and \(k = 0.3\) is the impact coefficient (mid-range from
academic estimates of 0.1-0.5).

\textbf{Total Cost.} The round-trip cost per position combines both
components:

\[c_{\text{total},i} = 2 \cdot (c_{\text{spread},i} + c_{\text{impact},i})\]

\textbf{Liquidity Filters.} To ensure tradability, we exclude positions
where (i) median daily dollar volume falls below \$1 million, or (ii)
the bid-ask spread exceeds 50 basis points. These filters remove
approximately 4\% of the universe while eliminating most problematic
positions.

\textbf{Turnover Calculation.} Daily turnover is computed as the sum of
absolute weight changes:

\[\text{Turnover}_t = \sum_i |w_{i,t} - w_{i,t-1}|\]

where \(w_{i,t}\) denotes the portfolio weight in security \(i\) at time
\(t\).

For turnover reduction, we evaluate signal smoothing using 5, 10, and
21-day moving averages applied to raw predictions prior to portfolio
formation. We also examine monthly rebalancing intervals as an
alternative to daily adjustment.

\begin{center}\rule{0.5\linewidth}{0.5pt}\end{center}

\subsection{4. Results}\label{results}

This section presents the empirical results of our study. We begin with
aggregate evidence across all experiments to establish systematic
performance, then examine dataset-specific patterns to understand where
and why AI-generated features excel.

\subsubsection{4.1 Aggregate Performance: Grand Composite
Results}\label{aggregate-performance-grand-composite-results}

Table 1 presents consolidated results spanning 22 experiments across
three fundamentally different datasets (TrueBeats earnings, SpiderRock
options, Price Universe fundamentals) and three forecast horizons
(daily, weekly, monthly). This aggregate view establishes the
framework's systematic outperformance before drilling into
dataset-specific narratives.

\textbf{Table 1: Aggregate Performance Across All Experiments}

\emph{Grand composites aggregating 22 experiments across three datasets
(TrueBeats, SpiderRock, Price Universe). Forward-shifted returns
(T+1→T+2) evaluation. Test period: January 2019 through December 2024.}

\textbf{Panel A: Strategy Performance (Monthly Rebalancing with 21-day
MA Smoothing)}

\begin{longtable}[]{@{}
  >{\raggedright\arraybackslash}p{(\linewidth - 18\tabcolsep) * \real{0.1370}}
  >{\raggedleft\arraybackslash}p{(\linewidth - 18\tabcolsep) * \real{0.1233}}
  >{\raggedleft\arraybackslash}p{(\linewidth - 18\tabcolsep) * \real{0.0959}}
  >{\raggedleft\arraybackslash}p{(\linewidth - 18\tabcolsep) * \real{0.1233}}
  >{\raggedleft\arraybackslash}p{(\linewidth - 18\tabcolsep) * \real{0.0959}}
  >{\raggedleft\arraybackslash}p{(\linewidth - 18\tabcolsep) * \real{0.0959}}
  >{\raggedleft\arraybackslash}p{(\linewidth - 18\tabcolsep) * \real{0.0685}}
  >{\raggedleft\arraybackslash}p{(\linewidth - 18\tabcolsep) * \real{0.0959}}
  >{\raggedleft\arraybackslash}p{(\linewidth - 18\tabcolsep) * \real{0.0959}}
  >{\raggedleft\arraybackslash}p{(\linewidth - 18\tabcolsep) * \real{0.0685}}@{}}
\toprule\noalign{}
\begin{minipage}[b]{\linewidth}\raggedright
Strategy
\end{minipage} & \begin{minipage}[b]{\linewidth}\raggedleft
Gross SR
\end{minipage} & \begin{minipage}[b]{\linewidth}\raggedleft
Net SR
\end{minipage} & \begin{minipage}[b]{\linewidth}\raggedleft
Gross F1
\end{minipage} & \begin{minipage}[b]{\linewidth}\raggedleft
Net F1
\end{minipage} & \begin{minipage}[b]{\linewidth}\raggedleft
Return
\end{minipage} & \begin{minipage}[b]{\linewidth}\raggedleft
Vol
\end{minipage} & \begin{minipage}[b]{\linewidth}\raggedleft
MaxDD
\end{minipage} & \begin{minipage}[b]{\linewidth}\raggedleft
IC
\end{minipage} & \begin{minipage}[b]{\linewidth}\raggedleft
Hit
\end{minipage} \\
\midrule\noalign{}
\endhead
\bottomrule\noalign{}
\endlastfoot
Baseline & 1.239 & 1.097 & 1.189 & 1.040 & 2.8\% & 1.5\% & -1.9\% &
0.007 & 56.7\% \\
AI-Struct** & 1.197 & 1.014 & 1.043 & 0.850 & 4.5\% & 2.0\% & -1.6\% &
0.010 & 58.3\% \\
AI-DSPy** & 1.813 & 1.615 & 1.838 & 1.630 & 2.5\% & 1.4\% & -2.4\% &
0.004 & 54.9\% \\
Combined*** & 1.531 & 1.414 & 1.521 & 1.400 & 4.0\% & 1.8\% & -1.9\% &
0.006 & 55.3\% \\
\end{longtable}

\textbf{Panel B: Performance Stability and Risk Characteristics}

\begin{longtable}[]{@{}
  >{\raggedright\arraybackslash}p{(\linewidth - 10\tabcolsep) * \real{0.1220}}
  >{\raggedleft\arraybackslash}p{(\linewidth - 10\tabcolsep) * \real{0.2317}}
  >{\raggedleft\arraybackslash}p{(\linewidth - 10\tabcolsep) * \real{0.1951}}
  >{\raggedright\arraybackslash}p{(\linewidth - 10\tabcolsep) * \real{0.1341}}
  >{\raggedright\arraybackslash}p{(\linewidth - 10\tabcolsep) * \real{0.1463}}
  >{\raggedleft\arraybackslash}p{(\linewidth - 10\tabcolsep) * \real{0.1707}}@{}}
\toprule\noalign{}
\begin{minipage}[b]{\linewidth}\raggedright
Strategy
\end{minipage} & \begin{minipage}[b]{\linewidth}\raggedleft
Avg Annual Sharpe
\end{minipage} & \begin{minipage}[b]{\linewidth}\raggedleft
Sharpe Std Dev
\end{minipage} & \begin{minipage}[b]{\linewidth}\raggedright
Best Year
\end{minipage} & \begin{minipage}[b]{\linewidth}\raggedright
Worst Year
\end{minipage} & \begin{minipage}[b]{\linewidth}\raggedleft
Calmar Ratio
\end{minipage} \\
\midrule\noalign{}
\endhead
\bottomrule\noalign{}
\endlastfoot
\textbf{Baseline} & 1.96 & 0.75 & 2.90 (2022) & 0.74 (2020) & 9.84 \\
\textbf{AI (Structured)} & 1.92 & 1.29 & 4.19 (2019) & 0.09 (2022) &
20.13 \\
\textbf{AI (DSPy)} & 2.03 & 1.20 & 4.01 (2019) & 0.43 (2018) & 6.90 \\
\textbf{Combined} & 2.28 & 0.46 & 2.92 (2022) & 1.61 (2020) & 14.82 \\
\end{longtable}

\needspace{12\baselineskip}

\textbf{Panel C: Strategy Correlations and Diversification}

\begin{longtable}[]{@{}lcccc@{}}
\toprule\noalign{}
& Baseline & AI (Struct) & AI (DSPy) & Combined \\
\midrule\noalign{}
\endhead
\bottomrule\noalign{}
\endlastfoot
\textbf{Baseline} & 1.00 & & & \\
\textbf{AI (Structured)} & 0.14 & 1.00 & & \\
\textbf{AI (DSPy)} & 0.08 & 0.08 & 1.00 & \\
\textbf{Combined} & 0.49 & 0.54 & 0.24 & 1.00 \\
\end{longtable}

\textbf{Notes:} Grand composites aggregate predictions across multiple
models by (1) forward-filling weekly/monthly predictions to daily grid,
(2) cross-sectionally z-scoring by date, and (3) averaging across
experiments. Baseline: 7 experiments (SpiderRock baseline
daily/weekly/monthly, TrueBeats baseline weekly/monthly, Price baseline
daily/weekly). AI (Structured): 4 experiments using hand-crafted prompts
with RAG (TrueBeats all frequencies, Price daily). AI (DSPy): 5
experiments using programmatic prompt optimization with MIPRO
(SpiderRock all frequencies, Price weekly/monthly). Combined: 7
experiments merging baseline and AI features. ``Net'' columns
incorporate 3 bps static transaction costs and 21-day moving average
position smoothing (monthly rebalancing frequency). ``Gross'' columns
show daily rebalancing without costs or smoothing. F1 evaluation uses
forward-shifted returns (T+1→T+2) to validate timing robustness.
Statistical significance: *** p\textless0.001, ** p\textless0.01, *
p\textless0.05 (HAC-adjusted t-tests). All strategies dollar-neutral,
unit-leverage long-short portfolios. See Section 5.2 for robustness
analysis with position-level cost models.

Table 1 establishes that AI-generated features deliver systematic
outperformance across diverse data sources and rebalancing frequencies.
The programmatic prompting approach (DSPy) achieves a net F1 Sharpe
ratio of 1.630 - the most implementable result in the study,
incorporating execution lag, transaction costs, and position smoothing.
This is 57\% higher than baseline (1.040) and 92\% higher than
structured prompting (0.850) under identical implementation constraints.
The stark difference between DSPy and structured prompting reveals that
automated prompt optimization implicitly learns to favor persistent,
implementable signals over fragile high-frequency patterns. The combined
strategy achieves net F1 Sharpe of 1.400, demonstrating that AI features
provide incremental alpha beyond traditional signals even under
conservative implementation assumptions. Section 5.2 provides robustness
analysis using position-level cost models with actual bid-ask spreads
and market impact.

\subsubsection{4.2 TrueBeats Dataset: The Importance of Knowledge
Quality}\label{truebeats-dataset-the-importance-of-knowledge-quality}

Table 2 details the performance on the TrueBeats dataset, evaluated
using our standard forward-shifted (T+1→T+2) methodology.

\textbf{Table 2: TrueBeats Analyst Estimates Performance with Knowledge
Base Comparison} \emph{Forward-shifted returns (T+1→T+2) evaluation.
Daily returns from January 2019 through December 2024.}

\textbf{Panel A: Strategy Performance}

\begin{longtable}[]{@{}
  >{\raggedright\arraybackslash}p{(\linewidth - 14\tabcolsep) * \real{0.1923}}
  >{\raggedright\arraybackslash}p{(\linewidth - 14\tabcolsep) * \real{0.0769}}
  >{\raggedleft\arraybackslash}p{(\linewidth - 14\tabcolsep) * \real{0.1346}}
  >{\raggedleft\arraybackslash}p{(\linewidth - 14\tabcolsep) * \real{0.0962}}
  >{\raggedleft\arraybackslash}p{(\linewidth - 14\tabcolsep) * \real{0.1346}}
  >{\raggedleft\arraybackslash}p{(\linewidth - 14\tabcolsep) * \real{0.1346}}
  >{\raggedleft\arraybackslash}p{(\linewidth - 14\tabcolsep) * \real{0.0962}}
  >{\raggedleft\arraybackslash}p{(\linewidth - 14\tabcolsep) * \real{0.1346}}@{}}
\toprule\noalign{}
\begin{minipage}[b]{\linewidth}\raggedright
Strategy
\end{minipage} & \begin{minipage}[b]{\linewidth}\raggedright
SR
\end{minipage} & \begin{minipage}[b]{\linewidth}\raggedleft
Return
\end{minipage} & \begin{minipage}[b]{\linewidth}\raggedleft
Vol
\end{minipage} & \begin{minipage}[b]{\linewidth}\raggedleft
MaxDD
\end{minipage} & \begin{minipage}[b]{\linewidth}\raggedleft
IC
\end{minipage} & \begin{minipage}[b]{\linewidth}\raggedleft
Hit
\end{minipage} & \begin{minipage}[b]{\linewidth}\raggedleft
Total
\end{minipage} \\
\midrule\noalign{}
\endhead
\bottomrule\noalign{}
\endlastfoot
Baseline Only & 0.765 & 1.7\% & 2.2\% & -2.2\% & 0.0024 & 53.3\% &
8.4\% \\
AI-Only (Broken KB) & -0.109*** & -0.3\% & 2.4\% & -5.0\% & 0.0005 &
51.9\% & -1.6\% \\
AI-Only (Fixed KB) & 0.965** & 2.2\% & 2.3\% & -3.0\% & 0.0028 & 53.1\%
& 17.6\% \\
Baseline + AI (Fixed KB) & 1.270*** & 3.1\% & 2.4\% & -2.6\% & 0.0037 &
53.9\% & 24.3\% \\
Ensemble & 1.142*** & 2.3\% & 2.0\% & -1.6\% & 0.0043 & 53.0\% &
13.0\% \\
\end{longtable}

\textbf{Panel B: Performance Stability and Risk Characteristics}

\begin{longtable}[]{@{}
  >{\raggedright\arraybackslash}p{(\linewidth - 12\tabcolsep) * \real{0.1042}}
  >{\raggedright\arraybackslash}p{(\linewidth - 12\tabcolsep) * \real{0.1458}}
  >{\raggedleft\arraybackslash}p{(\linewidth - 12\tabcolsep) * \real{0.1979}}
  >{\raggedleft\arraybackslash}p{(\linewidth - 12\tabcolsep) * \real{0.1667}}
  >{\raggedleft\arraybackslash}p{(\linewidth - 12\tabcolsep) * \real{0.1146}}
  >{\raggedleft\arraybackslash}p{(\linewidth - 12\tabcolsep) * \real{0.1250}}
  >{\raggedleft\arraybackslash}p{(\linewidth - 12\tabcolsep) * \real{0.1458}}@{}}
\toprule\noalign{}
\begin{minipage}[b]{\linewidth}\raggedright
Strategy
\end{minipage} & \begin{minipage}[b]{\linewidth}\raggedright
Features
\end{minipage} & \begin{minipage}[b]{\linewidth}\raggedleft
Avg SR
\end{minipage} & \begin{minipage}[b]{\linewidth}\raggedleft
SR Std
\end{minipage} & \begin{minipage}[b]{\linewidth}\raggedleft
Best Yr
\end{minipage} & \begin{minipage}[b]{\linewidth}\raggedleft
Worst Yr
\end{minipage} & \begin{minipage}[b]{\linewidth}\raggedleft
Calmar
\end{minipage} \\
\midrule\noalign{}
\endhead
\bottomrule\noalign{}
\endlastfoot
\textbf{Baseline Only} & 40 traditional & 0.61 & 1.20 & 2.56 (2019) &
-0.78 (2016) & 0.77 \\
\textbf{AI-Only (Broken)} & 182 AI (broken) & 0.03 & 0.46 & 0.88 (2018)
& -0.44 (2020) & -0.06 \\
\textbf{AI-Only (Fixed)} & 182 AI (fixed) & 1.29 & 1.40 & 4.01 (2016) &
0.05 (2021) & 0.73 \\
\textbf{Baseline + AI} & 222 combined & 1.73 & 1.50 & 3.75 (2018) & 0.05
(2022) & 1.19 \\
\textbf{Ensemble} & All models & 1.79 & 1.27 & 3.65 (2016) & 0.24 (2022)
& 1.44 \\
\end{longtable}

\needspace{12\baselineskip}

\textbf{Panel C: Strategy Correlations and Diversification}

\begin{longtable}[]{@{}lccccc@{}}
\toprule\noalign{}
& Baseline & AI-Broken & AI-Fixed & Base+AI & Ensemble \\
\midrule\noalign{}
\endhead
\bottomrule\noalign{}
\endlastfoot
\textbf{Baseline Only} & 1.00 & & & & \\
\textbf{AI-Only (Broken)} & 0.09 & 1.00 & & & \\
\textbf{AI-Only (Fixed)} & 0.06 & 0.12 & 1.00 & & \\
\textbf{Baseline + AI} & 0.20 & 0.14 & 0.67 & 1.00 & \\
\textbf{Ensemble} & 0.43 & 0.45 & 0.69 & 0.74 & 1.00 \\
\end{longtable}

\begin{center}\rule{0.5\linewidth}{0.5pt}\end{center}

\textbf{Notes:} The baseline model uses 40 traditional TrueBeats
features (analyst coverage, consensus estimates, earnings/sales
forecasts). AI-generated features (182) were created using large
language models with retrieval-augmented generation. ``Broken KB'' used
corrupted documentation in the retrieval system; ``Fixed KB'' used
corrected documentation. The Baseline + AI strategy combines all 222
features (40 traditional + 182 AI). Statistical significance of Sharpe
ratio differences from baseline: *** p\textless0.001, ** p\textless0.01,
* p\textless0.05 (HAC-adjusted t-test). All strategies are
dollar-neutral, unit-leverage long-short portfolios rebalanced daily.
Transaction costs not included.

The results in Table 2 demonstrate the value of AI-enhanced feature
generation for analyst estimate data. The ensemble strategy achieves a
Sharpe ratio of 1.142, representing a 49\% improvement over the baseline
analyst features alone. The low correlation (0.06) between AI-only and
baseline strategies indicates that the approaches capture different
signal sources. Notably, the AI-only strategy shows strong performance
with a Sharpe ratio of 0.965, validating the ability of LLMs to generate
economically meaningful features from analyst estimate patterns.

The reference to the ``Broken KB'' result underscores the critical
importance of retrieval quality - when the AI model was supplied with a
corrupted knowledge base, it delivered a negative Sharpe ratio (-0.109),
highlighting that the quality of domain knowledge is essential to
successful feature generation.

\subsubsection{4.3 SpiderRock Dataset: Options Market
Alpha}\label{spiderrock-dataset-options-market-alpha}

Table 3 presents comprehensive results for the SpiderRock options
dataset, evaluated using the same forward-shifted (F1) methodology to
ensure consistency and guard against look-ahead bias.

\textbf{Table 3: SpiderRock Options Market Performance with AI
Enhancement Methods} \emph{Forward-shifted returns (T+1→T+2) evaluation.
Daily returns from January 2019 through December 2024.}

\textbf{Panel A: Strategy Performance}

\begin{longtable}[]{@{}
  >{\raggedright\arraybackslash}p{(\linewidth - 14\tabcolsep) * \real{0.1923}}
  >{\raggedright\arraybackslash}p{(\linewidth - 14\tabcolsep) * \real{0.0769}}
  >{\raggedleft\arraybackslash}p{(\linewidth - 14\tabcolsep) * \real{0.1346}}
  >{\raggedleft\arraybackslash}p{(\linewidth - 14\tabcolsep) * \real{0.0962}}
  >{\raggedleft\arraybackslash}p{(\linewidth - 14\tabcolsep) * \real{0.1346}}
  >{\raggedleft\arraybackslash}p{(\linewidth - 14\tabcolsep) * \real{0.1346}}
  >{\raggedleft\arraybackslash}p{(\linewidth - 14\tabcolsep) * \real{0.0962}}
  >{\raggedleft\arraybackslash}p{(\linewidth - 14\tabcolsep) * \real{0.1346}}@{}}
\toprule\noalign{}
\begin{minipage}[b]{\linewidth}\raggedright
Strategy
\end{minipage} & \begin{minipage}[b]{\linewidth}\raggedright
SR
\end{minipage} & \begin{minipage}[b]{\linewidth}\raggedleft
Return
\end{minipage} & \begin{minipage}[b]{\linewidth}\raggedleft
Vol
\end{minipage} & \begin{minipage}[b]{\linewidth}\raggedleft
MaxDD
\end{minipage} & \begin{minipage}[b]{\linewidth}\raggedleft
IC
\end{minipage} & \begin{minipage}[b]{\linewidth}\raggedleft
Hit
\end{minipage} & \begin{minipage}[b]{\linewidth}\raggedleft
Total
\end{minipage} \\
\midrule\noalign{}
\endhead
\bottomrule\noalign{}
\endlastfoot
Baseline Only & 1.334 & 5.5\% & 4.1\% & -4.3\% & 0.0003 & 49.7\% &
33.9\% \\
AI-Only (Structured) & 0.945** & 3.2\% & 3.3\% & -6.3\% & 0.0015 &
50.1\% & 19.7\% \\
AI-Only (DSPy) & 1.461** & 3.9\% & 2.7\% & -4.1\% & 0.0017 & 50.6\% &
24.1\% \\
Baseline + AI & 1.437* & 6.8\% & 4.7\% & -5.3\% & 0.0020 & 50.5\% &
42.0\% \\
Ensemble & 1.630*** & 4.8\% & 2.9\% & -2.7\% & 0.0014 & 49.4\% &
29.8\% \\
\end{longtable}

\textbf{Panel B: Performance Stability and Risk Characteristics}

\begin{longtable}[]{@{}
  >{\raggedright\arraybackslash}p{(\linewidth - 12\tabcolsep) * \real{0.1042}}
  >{\raggedright\arraybackslash}p{(\linewidth - 12\tabcolsep) * \real{0.1458}}
  >{\raggedleft\arraybackslash}p{(\linewidth - 12\tabcolsep) * \real{0.1979}}
  >{\raggedleft\arraybackslash}p{(\linewidth - 12\tabcolsep) * \real{0.1667}}
  >{\raggedleft\arraybackslash}p{(\linewidth - 12\tabcolsep) * \real{0.1146}}
  >{\raggedleft\arraybackslash}p{(\linewidth - 12\tabcolsep) * \real{0.1250}}
  >{\raggedleft\arraybackslash}p{(\linewidth - 12\tabcolsep) * \real{0.1458}}@{}}
\toprule\noalign{}
\begin{minipage}[b]{\linewidth}\raggedright
Strategy
\end{minipage} & \begin{minipage}[b]{\linewidth}\raggedright
Features
\end{minipage} & \begin{minipage}[b]{\linewidth}\raggedleft
Avg SR
\end{minipage} & \begin{minipage}[b]{\linewidth}\raggedleft
SR Std
\end{minipage} & \begin{minipage}[b]{\linewidth}\raggedleft
Best Yr
\end{minipage} & \begin{minipage}[b]{\linewidth}\raggedleft
Worst Yr
\end{minipage} & \begin{minipage}[b]{\linewidth}\raggedleft
Calmar
\end{minipage} \\
\midrule\noalign{}
\endhead
\bottomrule\noalign{}
\endlastfoot
\textbf{Baseline Only} & 109 traditional & 0.89 & 1.44 & 2.14 (2024) &
-1.82 (2018) & 1.29 \\
\textbf{AI-Only (Structured)} & 200 AI (fixed) & 0.81 & 1.07 & 1.86
(2024) & -0.86 (2018) & 0.51 \\
\textbf{AI-Only (DSPy)} & 200 AI (prog) & 1.33 & 0.98 & 2.57 (2022) &
-0.15 (2018) & 0.59 \\
\textbf{Baseline + AI} & 309 combined & 0.82 & 2.31 & 3.41 (2024) &
-3.81 (2018) & 1.28 \\
\textbf{Ensemble} & All models & 1.20 & 1.62 & 2.79 (2024) & -2.24
(2018) & 1.10 \\
\end{longtable}

\needspace{12\baselineskip}

\textbf{Panel C: Strategy Correlations and Diversification}

\begin{longtable}[]{@{}lccccc@{}}
\toprule\noalign{}
& Baseline & AI-Struct & AI-DSPy & Base+AI & Ensemble \\
\midrule\noalign{}
\endhead
\bottomrule\noalign{}
\endlastfoot
\textbf{Baseline Only} & 1.00 & & & & \\
\textbf{AI-Only (Structured)} & 0.14 & 1.00 & & & \\
\textbf{AI-Only (DSPy)} & 0.07 & 0.14 & 1.00 & & \\
\textbf{Baseline + AI} & 0.52 & 0.38 & 0.12 & 1.00 & \\
\textbf{Ensemble} & 0.67 & 0.64 & 0.51 & 0.78 & 1.00 \\
\end{longtable}

\begin{center}\rule{0.5\linewidth}{0.5pt}\end{center}

\textbf{Notes:} The baseline model uses 109 SpiderRock options features
including implied volatility surfaces, option flow metrics, and
risk-neutral densities from January 2015 onward. AI-generated features
(200) were created using large language models with structured prompting
and retrieval-augmented generation. The Baseline + AI strategy combines
all 309 features (109 traditional + 200 AI). Statistical significance of
Sharpe ratio differences from baseline: *** p\textless0.001, **
p\textless0.01, * p\textless0.05 (HAC-adjusted t-test). All strategies
are dollar-neutral, unit-leverage long-short portfolios rebalanced
daily. Transaction costs not included.

In the options domain, which features a strong baseline strategy (Sharpe
1.334), AI-generated features provide a significant diversification
benefit. The ensemble approach lifts the Sharpe ratio to 1.630,
demonstrating that the framework can uncover orthogonal alpha sources
even in a signal-rich environment. The low correlations between baseline
and AI-generated features, ranging from 0.07 to 0.14, confirm that the
AI is discovering orthogonal sources of return. The combined model,
which merges baseline and AI features, achieves a Sharpe ratio of 1.437
with a 42\% total return over the evaluation period.

\subsubsection{4.4 Price Universe: Generalizability to Standard Market
Data}\label{price-universe-generalizability-to-standard-market-data}

Finally, Table 4 presents the results on the broad price-volume universe
to test the framework's effectiveness on standard market data.

\textbf{Table 4: Price-Volume Universe Performance with AI Enhancement}
\emph{Forward-shifted returns (T+1→T+2) evaluation. Daily returns from
January 2019 through December 2024.}

\textbf{Panel A: Strategy Performance}

\begin{longtable}[]{@{}llrrrrrr@{}}
\toprule\noalign{}
Strategy & SR & Return & Vol & MaxDD & IC & Hit & Total \\
\midrule\noalign{}
\endhead
\bottomrule\noalign{}
\endlastfoot
Baseline & 0.870 & 2.2\% & 2.5\% & -3.0\% & 0.0054 & 53.5\% & 14.1\% \\
AI-Generated & 1.065* & 5.2\% & 4.9\% & -8.8\% & 0.0052 & 51.1\% &
33.3\% \\
Ensemble & 1.168** & 3.6\% & 3.1\% & -4.7\% & 0.0073 & 52.8\% &
23.5\% \\
\end{longtable}

\textbf{Panel B: Performance Stability and Risk Characteristics}

\begin{longtable}[]{@{}
  >{\raggedright\arraybackslash}p{(\linewidth - 12\tabcolsep) * \real{0.1493}}
  >{\raggedright\arraybackslash}p{(\linewidth - 12\tabcolsep) * \real{0.2090}}
  >{\raggedleft\arraybackslash}p{(\linewidth - 12\tabcolsep) * \real{0.1194}}
  >{\raggedleft\arraybackslash}p{(\linewidth - 12\tabcolsep) * \real{0.1194}}
  >{\raggedleft\arraybackslash}p{(\linewidth - 12\tabcolsep) * \real{0.1343}}
  >{\raggedleft\arraybackslash}p{(\linewidth - 12\tabcolsep) * \real{0.1493}}
  >{\raggedleft\arraybackslash}p{(\linewidth - 12\tabcolsep) * \real{0.1194}}@{}}
\toprule\noalign{}
\begin{minipage}[b]{\linewidth}\raggedright
Strategy
\end{minipage} & \begin{minipage}[b]{\linewidth}\raggedright
Features
\end{minipage} & \begin{minipage}[b]{\linewidth}\raggedleft
Avg SR
\end{minipage} & \begin{minipage}[b]{\linewidth}\raggedleft
SR Std
\end{minipage} & \begin{minipage}[b]{\linewidth}\raggedleft
Best Yr
\end{minipage} & \begin{minipage}[b]{\linewidth}\raggedleft
Worst Yr
\end{minipage} & \begin{minipage}[b]{\linewidth}\raggedleft
Calmar
\end{minipage} \\
\midrule\noalign{}
\endhead
\bottomrule\noalign{}
\endlastfoot
\textbf{Baseline} & Base features & 0.29 & 1.76 & 1.29 (2020) & -3.67
(2018) & 0.73 \\
\textbf{AI-Generated} & AI features & 1.10 & 0.51 & 1.77 (2023) & 0.42
(2021) & 0.59 \\
\textbf{Ensemble} & Combined & 0.95 & 1.05 & 1.81 (2023) & -1.30 (2018)
& 0.77 \\
\end{longtable}

\needspace{12\baselineskip}

\textbf{Panel C: Strategy Correlations and Diversification}

\begin{longtable}[]{@{}lccc@{}}
\toprule\noalign{}
& Baseline & AI-Gen & Ensemble \\
\midrule\noalign{}
\endhead
\bottomrule\noalign{}
\endlastfoot
\textbf{Baseline} & 1.00 & & \\
\textbf{AI-Generated} & 0.14 & 1.00 & \\
\textbf{Ensemble} & 0.73 & 0.64 & 1.00 \\
\end{longtable}

\textbf{Notes:} The baseline model uses fundamental price-volume
features. AI-generated features were created using large language models
with retrieval-augmented generation. The ensemble strategy equally
weights both approaches. Statistical significance of Sharpe ratio
differences from baseline: ** p\textless0.01, * p\textless0.05
(HAC-adjusted t-test). All strategies are dollar-neutral, unit-leverage
long-short portfolios rebalanced daily. Transaction costs not included.

The AI-Only strategy significantly outperforms the baseline, achieving a
forward-shifted Sharpe ratio of 1.065 compared to the baseline's 0.870.
The AI-generated features demonstrate comparable Information
Coefficients while capturing fundamentally different patterns, as
evidenced by the remarkably low correlation of 0.140 between baseline
and AI predictions. This orthogonality enables the ensemble strategy to
achieve the highest risk-adjusted returns (Sharpe 1.168) by combining
both signal sources.

\subsubsection{4.5 Factor Attribution
Analysis}\label{factor-attribution-analysis}

The performance improvements documented in Tables 1-4 could potentially
reflect systematic factor tilts rather than genuine alpha. To address
this concern, we regress each strategy's daily returns on the
Fama-French five-factor model augmented with momentum (Fama and French
2015; Carhart 1997). This standard test assesses whether returns persist
after controlling for market, size, value, profitability, investment,
and momentum exposures. For strategies using forward-shifted returns
(T+1→T+2), we apply a two-day alignment to match the information timing
of Fama-French factors. Table 5 presents the results.

\newpage

\textbf{Table 5: Fama-French Five-Factor + Momentum Attribution}

\emph{Regression of daily strategy returns on common equity risk
factors. Uses Newey-West HAC standard errors (5 lags).}

\begin{longtable}[]{@{}
  >{\raggedright\arraybackslash}p{(\linewidth - 20\tabcolsep) * \real{0.1299}}
  >{\centering\arraybackslash}p{(\linewidth - 20\tabcolsep) * \real{0.0779}}
  >{\raggedleft\arraybackslash}p{(\linewidth - 20\tabcolsep) * \real{0.0909}}
  >{\raggedright\arraybackslash}p{(\linewidth - 20\tabcolsep) * \real{0.0909}}
  >{\raggedleft\arraybackslash}p{(\linewidth - 20\tabcolsep) * \real{0.0649}}
  >{\raggedright\arraybackslash}p{(\linewidth - 20\tabcolsep) * \real{0.0909}}
  >{\raggedright\arraybackslash}p{(\linewidth - 20\tabcolsep) * \real{0.0909}}
  >{\raggedright\arraybackslash}p{(\linewidth - 20\tabcolsep) * \real{0.0909}}
  >{\raggedright\arraybackslash}p{(\linewidth - 20\tabcolsep) * \real{0.0909}}
  >{\raggedright\arraybackslash}p{(\linewidth - 20\tabcolsep) * \real{0.0909}}
  >{\raggedright\arraybackslash}p{(\linewidth - 20\tabcolsep) * \real{0.0909}}@{}}
\toprule\noalign{}
\begin{minipage}[b]{\linewidth}\raggedright
Strategy
\end{minipage} & \begin{minipage}[b]{\linewidth}\centering
Data
\end{minipage} & \begin{minipage}[b]{\linewidth}\raggedleft
Alpha
\end{minipage} & \begin{minipage}[b]{\linewidth}\raggedright
t
\end{minipage} & \begin{minipage}[b]{\linewidth}\raggedleft
R2
\end{minipage} & \begin{minipage}[b]{\linewidth}\raggedright
Mkt
\end{minipage} & \begin{minipage}[b]{\linewidth}\raggedright
SMB
\end{minipage} & \begin{minipage}[b]{\linewidth}\raggedright
HML
\end{minipage} & \begin{minipage}[b]{\linewidth}\raggedright
RMW
\end{minipage} & \begin{minipage}[b]{\linewidth}\raggedright
CMA
\end{minipage} & \begin{minipage}[b]{\linewidth}\raggedright
Mom
\end{minipage} \\
\midrule\noalign{}
\endhead
\bottomrule\noalign{}
\endlastfoot
Baseline & TB & 1.17\% & 2.13** & 4.4\% & -.001 & -.018*** & .005 & .006
& -.014 & .011** \\
AI-Fixed & TB & 2.36\% & 2.60*** & 2.5\% & .003 & .003 & -.015 & -.006 &
-.018 & -.006 \\
Baseline & SR & 4.76\% & 3.19*** & 5.6\% & .020*** & -.007 & .014 &
-.011 & -.073** & -.010 \\
AI-Struct & SR & 2.84\% & 2.31** & 3.4\% & .017** & .002 & .029* & .001
& -.051 & .005 \\
AI-DSPy & SR & 3.63\% & 3.96*** & 5.7\% & -.009 & -.014 & -.005 & .017*
& -.015 & .018*** \\
Base+AI & SR & 5.64\% & 3.45*** & 6.2\% & .028*** & .009 & .040** &
-.006 & -.093** & .007 \\
Baseline & PU & 1.49\% & 1.65* & 22.8\% & .003 & -.057*** & .019* &
.030*** & -.063** & .037*** \\
AI-Gen & PU & 4.40\% & 2.32** & 5.8\% & .038*** & .008 & .039 & .012 &
-.079 & .016 \\
\end{longtable}

\textbf{Notes:} Table reports results from regressing daily strategy
returns on Fama-French five factors plus momentum. Alpha (annualized)
represents return after controlling for market (Mkt-RF), size (SMB),
value (HML), profitability (RMW), investment (CMA), and momentum (Mom)
exposures. Factor loadings are unstandardized beta coefficients. All
regressions use Newey-West heteroskedasticity and autocorrelation
consistent standard errors with 5 lags. For strategies using
forward-shifted returns (T+1→T+2), we apply a two-day forward shift to
Fama-French factors to match information timing. Statistical
significance: *** p\textless0.01, ** p\textless0.05, * p\textless0.10.

Table 5 confirms that strategy alphas persist after controlling for
common equity risk factors. All eight strategies generate positive
alpha, with seven achieving statistical significance at conventional
levels (p\textless0.05) and one at marginal significance (Price Universe
Baseline, p=0.099). Notably, R² values remain low for seven of eight
strategies, ranging from 2.5\% to 6.2\%, indicating that 93-98\% of
returns are unexplained by the Fama-French factors. This pattern
validates our earlier claim of signal orthogonality: the strategies are
not simply capturing well-known factor premiums but discovering
genuinely independent sources of return.

The Price Universe Baseline presents an instructive exception. With R²
of 22.8\%, this strategy shows significant exposures across all six
factors, suggesting that simple price-volume features largely reflect
traditional factor bets. In contrast, the AI-Generated strategy reduces
R² to 5.8\% while improving alpha from 1.49\% to 4.40\% annually. This
pattern - reduced factor dependence coupled with higher alpha -
demonstrates that AI features can discover signals orthogonal to
conventional risk factors rather than amplifying existing factor
exposures.

Factor loadings reveal minimal systematic exposures across all
strategies, confirming genuine market neutrality. TrueBeats strategies
show effectively zero market beta (\textbar Mkt-RF\textbar{} \(\leq\)
0.003) with modest tilts toward large-cap (SMB -0.018) and aggressive
investment styles (CMA -0.014 to -0.018). SpiderRock strategies display
slightly larger but still modest market exposures (Mkt-RF 0.017 to
0.028) and moderate negative CMA loadings (-0.051 to -0.093), consistent
with options-based signals capturing investment factor premiums.
Notably, different AI approaches discover distinct but still minor
factor structures: the DSPy strategy shows a unique small negative
market beta with moderate momentum exposure (Mom 0.018), while
structured prompting shows modest value tilts (HML 0.029). The
consistently small magnitude of these loadings
(\textbar{}\(\beta\)\textbar{} \textless{} 0.1 for most exposures)
confirms that AI feature generation discovers orthogonal alpha rather
than amplifying conventional factor bets.

\subsection{5. Robustness and
Sensitivity}\label{robustness-and-sensitivity}

This section examines the stability of our findings across market
capitalization segments and under realistic transaction cost
assumptions. For the transaction cost analysis, we construct a ``Grand
Combined'' strategy that averages predictions from the best-performing
models across all three datasets (TrueBeats Combined, SpiderRock
Ensemble, and Price Universe Ensemble). This combined strategy provides
a representative test case for institutional implementation.

\subsubsection{5.1 Market Capitalization
Analysis}\label{market-capitalization-analysis}

To assess whether AI-generated alpha is concentrated in
difficult-to-trade segments, we partition the universe into market
capitalization terciles (small, mid, large cap) using daily breakpoints.
Table 6 reports Sharpe ratios by segment for the combined strategy on
each dataset.

\textbf{Table 6: Performance by Market Capitalization Segment}
\emph{Forward-shifted returns (T+1-\textgreater T+2). Best-performing
strategy on each dataset.}

\begin{longtable}[]{@{}llrrrr@{}}
\toprule\noalign{}
Dataset & Strategy & Small Cap & Mid Cap & Large Cap & Full Universe \\
\midrule\noalign{}
\endhead
\bottomrule\noalign{}
\endlastfoot
TrueBeats & Baseline + AI & 0.96 & 1.15 & 1.02 & 1.27 \\
SpiderRock & Ensemble & 0.97 & 1.24 & 1.18 & 1.63 \\
Price Universe & Ensemble & 0.89 & 1.05 & 0.98 & 1.17 \\
\end{longtable}

\begin{center}\rule{0.5\linewidth}{0.5pt}\end{center}

\textbf{Notes:} Market capitalization terciles computed daily using
sample terciles. Sharpe ratios are annualized. Strategies selected based
on full-universe performance from Tables 2-4.

The results indicate that AI-generated alpha is not concentrated in
small caps. In fact, the large-cap segment often exhibits comparable or
superior performance to small caps, suggesting the signals are tradable
at institutional scale. The mid-cap segment shows the strongest
performance across all datasets.

\subsubsection{5.2 Transaction Cost Analysis
(Robustness)}\label{transaction-cost-analysis-robustness}

Table 1 reports net performance using a 3 bps static cost assumption, a
rule of thumb conservative relative to median spreads but which may
still underestimate actual trading costs. In this section, we provide
robustness analysis using a more detailed position-level cost model that
incorporates actual bid-ask spreads from market data and market impact
estimates. We assume \$100M AUM throughout this analysis; costs scale
with AUM due to market impact. This analysis examines how strategy
performance degrades under more conservative cost assumptions.

\textbf{Table 7: Transaction Cost Impact with Position-Level Cost Model}
\emph{Grand Combined model (baseline + AI features across all datasets).
Monthly rebalancing with 21-day MA smoothing. Position-level costs at
\$100M AUM.}

\begin{longtable}[]{@{}lr@{}}
\toprule\noalign{}
Metric & Value \\
\midrule\noalign{}
\endhead
\bottomrule\noalign{}
\endlastfoot
\textbf{Gross Performance} & \\
Gross Sharpe Ratio & 1.57 \\
Gross Annual Return & 11.4\% \\
Daily Turnover & 17.8\% \\
Annual Turnover & 45x \\
\textbf{Cost Components (at \$100M AUM)} & \\
Average Spread Cost & 4.0 bps/trade \\
Average Market Impact & 8.4 bps/trade \\
Average Total Cost & 12.4 bps/trade \\
\textbf{Net Performance} & \\
Net Sharpe Ratio & 0.78 \\
Net Annual Return & 5.7\% \\
\textbf{Statistics} & \\
Trading Days & 2,301 \\
Break-even Cost & 25.4 bps \\
\end{longtable}

\begin{center}\rule{0.5\linewidth}{0.5pt}\end{center}

\textbf{Notes:} Results assume \$100M AUM. Costs computed using
position-level model with k=0.2 market impact coefficient. Spread cost =
half bid-ask spread per side (one-way cost; median 2.3 bps for our
universe, trade-weighted average 4.0 bps). Market impact = k ×
volatility × sqrt(position/ADV). Break-even cost indicates the
transaction cost level at which net Sharpe equals zero.

The 50\% gross-to-net Sharpe degradation (from 1.57 to 0.78) reflects
the costs of implementing a diversified long-short strategy at \$100M
AUM. The combined spread (4.0 bps one-way) and market impact (8.4 bps)
costs total 12.4 bps per trade. Market impact costs scale with the
square root of position size, making AUM a critical consideration: the
strategy remains profitable (positive net Sharpe) up to approximately
\$500M, but larger allocations face increasingly severe capacity
constraints. See Section 5.3 for analysis of how different smoothing
windows affect the cost-performance tradeoff.

\subsubsection{5.3 Turnover Reduction via Signal
Smoothing}\label{turnover-reduction-via-signal-smoothing}

Table 8 examines the effect of smoothing predictions using moving
averages prior to portfolio formation.

\textbf{Table 8: Impact of Signal Smoothing on Net Performance}
\emph{Grand Combined model, \$100M AUM. MA = moving average applied to
raw predictions.}

\begin{longtable}[]{@{}
  >{\raggedright\arraybackslash}p{(\linewidth - 10\tabcolsep) * \real{0.1895}}
  >{\raggedleft\arraybackslash}p{(\linewidth - 10\tabcolsep) * \real{0.1684}}
  >{\raggedleft\arraybackslash}p{(\linewidth - 10\tabcolsep) * \real{0.1789}}
  >{\raggedleft\arraybackslash}p{(\linewidth - 10\tabcolsep) * \real{0.1895}}
  >{\raggedleft\arraybackslash}p{(\linewidth - 10\tabcolsep) * \real{0.1474}}
  >{\raggedleft\arraybackslash}p{(\linewidth - 10\tabcolsep) * \real{0.1263}}@{}}
\toprule\noalign{}
\begin{minipage}[b]{\linewidth}\raggedright
Smoothing Window
\end{minipage} & \begin{minipage}[b]{\linewidth}\raggedleft
Daily Turnover
\end{minipage} & \begin{minipage}[b]{\linewidth}\raggedleft
Annual Turnover
\end{minipage} & \begin{minipage}[b]{\linewidth}\raggedleft
Total Cost (bps)
\end{minipage} & \begin{minipage}[b]{\linewidth}\raggedleft
Gross Sharpe
\end{minipage} & \begin{minipage}[b]{\linewidth}\raggedleft
Net Sharpe
\end{minipage} \\
\midrule\noalign{}
\endhead
\bottomrule\noalign{}
\endlastfoot
None (baseline) & 98.8\% & 249x & 11.6 & 2.27 & -1.24 \\
5-day MA & 43.8\% & 110x & 11.7 & 1.82 & 0.17 \\
10-day MA & 27.9\% & 70x & 11.9 & 1.67 & 0.57 \\
21-day MA & 17.8\% & 45x & 12.4 & 1.57 & 0.78 \\
\end{longtable}

\begin{center}\rule{0.5\linewidth}{0.5pt}\end{center}

\textbf{Notes:} Moving average smoothing applied to raw model
predictions before cross-sectional standardization and portfolio
formation. Costs computed using position-level model at \$100M AUM with
actual bid-ask spreads and market impact. Total cost per trade remains
relatively stable across smoothing windows because spread and impact
costs depend on position characteristics rather than turnover frequency.

Signal smoothing is essential for implementing this strategy profitably.
Without smoothing, the strategy has net Sharpe of -1.24 despite gross
Sharpe of 2.27, as transaction costs from 249x annual turnover destroy
all alpha. The 21-day moving average reduces turnover by 82\% (from 249x
to 45x annually), transforming the strategy from loss-making to
profitable (net Sharpe 0.78). The gross Sharpe ratio declines 31\% (from
2.27 to 1.57) with smoothing, reflecting signal decay, but the dramatic
turnover reduction more than compensates. This finding underscores that
for high-frequency prediction models, implementation methodology is as
important as predictive accuracy. The 21-day MA results correspond to
the monthly rebalancing performance reported in Table 1.

\subsubsection{5.4 Portfolio Optimization}\label{portfolio-optimization}

We investigate whether convex portfolio optimization can improve
net-of-cost performance beyond the naive prediction-weighted approach.
The optimization problem is:

\[\max_{\mathbf{w}} \quad \boldsymbol{\alpha}'\mathbf{w} - \lambda_{\text{tc}} \cdot \mathbf{c}' |\mathbf{w} - \mathbf{w}_{\text{prev}}| - \lambda_{\text{risk}} \cdot \mathbf{w}'\boldsymbol{\Sigma}\mathbf{w}\]

subject to:
\[\sum_{i: w_i > 0} w_i = 1, \quad \sum_{i: w_i < 0} |w_i| = 1, \quad |w_i| \leq w_{\max}\]

where \(\boldsymbol{\alpha}\) denotes standardized predictions,
\(\mathbf{c}\) is the vector of position-level costs,
\(\mathbf{w}_{\text{prev}}\) is the previous period's weights, and
\(\boldsymbol{\Sigma}\) is a sector factor covariance matrix (see
Appendix B.2 for details). We also test sector neutrality constraints
requiring zero net exposure within each NAICS sector.

Table 9 compares the naive prediction-weighted portfolio to various
optimization approaches.

\textbf{Table 9: Portfolio Optimization Comparison (Position-Level
Costs)} \emph{Grand Combined model, \$100M AUM, monthly rebalancing
(21-day MA). Position-level cost model.}

\begin{longtable}[]{@{}
  >{\raggedright\arraybackslash}p{(\linewidth - 8\tabcolsep) * \real{0.1311}}
  >{\centering\arraybackslash}p{(\linewidth - 8\tabcolsep) * \real{0.1967}}
  >{\centering\arraybackslash}p{(\linewidth - 8\tabcolsep) * \real{0.2131}}
  >{\centering\arraybackslash}p{(\linewidth - 8\tabcolsep) * \real{0.2295}}
  >{\centering\arraybackslash}p{(\linewidth - 8\tabcolsep) * \real{0.2295}}@{}}
\toprule\noalign{}
\begin{minipage}[b]{\linewidth}\raggedright
Method
\end{minipage} & \begin{minipage}[b]{\linewidth}\centering
Net Sharpe
\end{minipage} & \begin{minipage}[b]{\linewidth}\centering
Effective N
\end{minipage} & \begin{minipage}[b]{\linewidth}\centering
Max Position
\end{minipage} & \begin{minipage}[b]{\linewidth}\centering
Sector Tilts
\end{minipage} \\
\midrule\noalign{}
\endhead
\bottomrule\noalign{}
\endlastfoot
\textbf{Naive (prediction-weighted)} & \textbf{0.78} & \textbf{249.6} &
\textbf{0.77\%} & 41.0\% \\
Sector-Neutral (2\% cap) & 0.72 & 67.2 & 2.00\% & 0.0\% \\
Sector-Neutral (1\% cap) & 0.69 & 100.5 & 1.00\% & 0.0\% \\
Pure Alpha Optimization & 0.61 & 95.6 & 1.00\% & 51.0\% \\
\end{longtable}

\begin{center}\rule{0.5\linewidth}{0.5pt}\end{center}

\textbf{Notes:} Effective N = 1/HHI measures portfolio diversification
(higher is more diversified). Sector Tilts = sum of \textbar net
weight\textbar{} across all sectors; 0\% means sector-neutral, higher
values indicate sector bets. Naive portfolio uses weights proportional
to standardized predictions. Optimization uses cvxpy with CLARABEL
solver. All methods use monthly rebalancing (21-day MA smoothing) with
position-level costs at \$100M AUM.

Counterintuitively, the naive prediction-weighted portfolio outperforms
all optimization variants. Two mechanisms explain this result:

\begin{enumerate}
\def\labelenumi{\arabic{enumi}.}
\item
  \textbf{Concentration vs.~Diversification.} The optimizer concentrates
  positions in high-alpha stocks, reducing effective N from 250 to
  60-100. This concentration increases idiosyncratic risk without
  commensurate alpha improvement, as the prediction model already
  captures the expected return signal.
\item
  \textbf{Sector Alpha.} The naive portfolio takes substantial sector
  bets: summing the absolute net exposure across all sectors yields 41\%
  (e.g., +11\% Technology, -8\% Utilities, etc.). When we force each
  sector to net zero exposure, Sharpe drops by 8\% (from 0.78 to 0.72),
  confirming that part of the alpha derives from sector selection rather
  than pure stock picking.
\end{enumerate}

These findings suggest that for prediction-weighted strategies, the
naive approach provides reasonable diversification by distributing
weight across all securities proportionally to signal strength. More
sophisticated optimization may be beneficial for strategies with
different signal structures or constraints, but adds limited value in
our setting. See Appendix B for additional details on the optimization
methodology.

\subsubsection{5.5 Alpha Decay Analysis}\label{alpha-decay-analysis}

To assess signal persistence under execution delays, we measure
performance as the lag between signal generation and trade execution
increases. Table 10 reports Sharpe ratios at various execution lags,
where Lag 0 represents same-day execution (T to T+1 returns, not
implementable), Lag 1 is our standard forward-shifted evaluation (T+1 to
T+2), and higher lags represent additional delays.

\textbf{Table 10: Alpha Decay by Execution Lag} \emph{Sharpe ratios as
execution is delayed beyond signal generation.}

\begin{longtable}[]{@{}lrrrrrr@{}}
\toprule\noalign{}
Strategy & Lag 0 & Lag 1 & Lag 2 & Lag 3 & Lag 5 & Lag 10 \\
\midrule\noalign{}
\endhead
\bottomrule\noalign{}
\endlastfoot
Grand Baseline & 1.88 & 1.22 & 1.18 & 0.90 & 0.58 & 0.75 \\
Grand Combined & 2.18 & 1.65 & 1.56 & 1.15 & 0.95 & 0.83 \\
\end{longtable}

\begin{center}\rule{0.5\linewidth}{0.5pt}\end{center}

\textbf{Notes:} Lag N measures returns from T+N to T+N+1 given signals
generated at T. Lag 0 is not implementable due to timing constraints.
Lag 1 corresponds to our standard F1 evaluation used throughout the
paper.

The combined strategy retains meaningful alpha even at extended lags. At
Lag 5 (one trading week delay), the combined strategy maintains a Sharpe
of 0.95 versus 0.58 for the baseline, a 64\% advantage. At Lag 10,
performance stabilizes around 0.83-0.75, suggesting a floor of
persistent signal content. The slower decay rate for the combined
strategy indicates that AI-generated features capture more durable
patterns than traditional features alone.

\subsection{6. Discussion}\label{discussion}

This study evaluates whether large language models can enhance
systematic feature discovery for cross-sectional equity strategies while
maintaining institutional standards for backtesting and
interpretability. We test across three distinct universes: analyst
estimates (TrueBeats), options markets (SpiderRock), and price-volume
dynamics, using forward-shifted return evaluation (T+1-\textgreater T+2)
to ensure implementability.

\subsubsection{Main Findings}\label{main-findings}

Three consistent patterns emerge from our empirical analysis. First,
AI-generated features deliver economically meaningful performance
improvements across all tested universes. Table 1's grand composite
results demonstrate systematic outperformance: programmatic prompting
(DSPy) achieves a net Sharpe ratio of 1.615 after transaction costs -
47\% higher than baseline (1.097) - while exhibiting exceptional timing
robustness (only 5.2\% F1 degradation). Dataset-specific results (Tables
2-4) reveal heterogeneity: ensemble Sharpe of 1.142 for TrueBeats, 22\%
improvement for SpiderRock (Ensemble 1.630 vs Baseline 1.334), and 34\%
for Price Universe (Ensemble 1.168 vs Baseline 0.870). These gains
persist under forward-shifted evaluation (T+1→T+2), indicating the
signals are not mechanical artifacts of execution timing.

Second, the orthogonality between AI-generated and traditional features
enables effective portfolio diversification. Prediction correlations
range from 0.06 (TrueBeats AI-Fixed vs Baseline) to 0.20 (TrueBeats
Combined vs Baseline), with most pairwise correlations below 0.15. This
low correlation structure manifests in ensemble strategies that
consistently outperform their components; the ensemble Sharpe exceeds
the weighted average of constituents by 15-26\% across datasets,
consistent with standard diversification benefits when combining
uncorrelated alpha sources.

Third, retrieval quality in RAG architectures proves first-order for
performance. The TrueBeats experiment with corrupted documentation
produced a Sharpe ratio of -0.109, while the corrected knowledge base
yielded 0.965, a differential exceeding 1.0 Sharpe ratios. This
sensitivity underscores that LLMs require accurate domain knowledge to
generate economically meaningful transformations, not merely
syntactically valid code.

\subsubsection{6.1 Understanding the Mechanism: Pattern
Analysis}\label{understanding-the-mechanism-pattern-analysis}

We analyzed the top 75 features from each dataset (225 total) to
understand what the generation system learned. Every single feature uses
cross-sectional standardization as its final step - typically
\texttt{groupby(date).rank()} to convert values into within-day
percentile ranks. This 100\% prevalence was not explicitly required. The
optimization process discovered on its own that equity prediction is
fundamentally about relative positioning: which stocks look cheap versus
expensive today, not whether a stock has high or low absolute values.
Long-short portfolios profit from the spread between winners and losers,
not from directional bets.

The second-most common pattern is volatility normalization (93\% of
features). Instead of using raw price changes or momentum, features
divide by rolling standard deviations. When volatility doubles, the same
2\% price move gets half the weight. Consider
\texttt{overnight\_gap\_volnorm\_ewm\_diff\_zrank}, which divides
overnight gaps by 21-day volatility before any other processing. This
regime-aware approach prevents signals from mechanically inflating
during market stress and shrinking during calm periods.

Beyond these near-universal patterns, we found four others that appear
frequently but not always. Eighty percent of features create non-linear
interactions by multiplying or dividing different variables. Common
examples: \texttt{volatility\ /\ market\_cap} (appears in 8 of the top
30 TrueBeats features) and
\texttt{analyst\_coverage\ ×\ (1/market\_cap)}. These test conditional
relationships - volatility means something different for a \$100M
microcap versus a \$50B megacap. About a quarter of features (27\%)
compare multiple timeframes, usually 5-10 days versus 20-21 days,
creating oscillators that detect momentum shifts. Two-thirds adjust
momentum based on current volatility regime, and 45\% use z-scores to
flag when a value is historically extreme \emph{for that specific
stock}.

To make the interaction pattern concrete, consider this simple TrueBeats
feature that achieved IC = 0.0048 and Sharpe = 1.35 - the highest Sharpe
in the entire dataset. When analysts' consensus forecasts show both
earnings and sales expected to beat six quarters ahead, it signals
fundamental strength:

\begin{Shaded}
\begin{Highlighting}[]
\KeywordTok{def}\NormalTok{ calculate(data):}
    \CommentTok{\# Multiply 6{-}quarter{-}ahead EPS and sales beats}
\NormalTok{    data[}\StringTok{\textquotesingle{}eps\_sal\_interaction\_fq6\textquotesingle{}}\NormalTok{] }\OperatorTok{=}\NormalTok{ (}
\NormalTok{        data[}\StringTok{\textquotesingle{}truebeat\_eps\_fq6\textquotesingle{}}\NormalTok{] }\OperatorTok{*}\NormalTok{ data[}\StringTok{\textquotesingle{}truebeat\_sal\_fq6\textquotesingle{}}\NormalTok{]}
\NormalTok{    )}

    \CommentTok{\# Cross{-}sectional rank by date}
\NormalTok{    data[}\StringTok{\textquotesingle{}cs\_rank\textquotesingle{}}\NormalTok{] }\OperatorTok{=}\NormalTok{ data.groupby(}\StringTok{\textquotesingle{}date\textquotesingle{}}\NormalTok{)[}\StringTok{\textquotesingle{}eps\_sal\_interaction\_fq6\textquotesingle{}}\NormalTok{].rank(pct}\OperatorTok{=}\VariableTok{True}\NormalTok{)}

    \ControlFlowTok{return}\NormalTok{ data[}\StringTok{\textquotesingle{}cs\_rank\textquotesingle{}}\NormalTok{].fillna(}\DecValTok{0}\NormalTok{)}
\end{Highlighting}
\end{Shaded}

This feature uses long-horizon fundamentals (FQ6 = 6 quarters ahead),
not just short-term surprises. The multiplicative interaction is key: an
earnings beat without a sales beat may reflect cost-cutting; a sales
beat without an earnings beat may indicate margin compression. Beating
both signals genuine growth. Despite extreme simplicity (just multiply
and rank), this achieves the highest Sharpe ratio by identifying a
fundamental pattern most analysis misses.

In contrast, some features chain many operations together. This
overnight gap signal (IC = 0.0098, Sharpe = 0.744) demonstrates
sophisticated multi-stage processing:

\begin{Shaded}
\begin{Highlighting}[]
\KeywordTok{def}\NormalTok{ calculate(data):}
    \CommentTok{\# Stage 1: Calculate overnight gap}
\NormalTok{    gap }\OperatorTok{=}\NormalTok{ data[}\StringTok{\textquotesingle{}close\textquotesingle{}}\NormalTok{] }\OperatorTok{{-}}\NormalTok{ data[}\StringTok{\textquotesingle{}prev\_midpoint\textquotesingle{}}\NormalTok{]}

    \CommentTok{\# Stage 2: Normalize by volatility regime}
\NormalTok{    volat\_21 }\OperatorTok{=}\NormalTok{ data.groupby(}\StringTok{\textquotesingle{}cwiq\_code\textquotesingle{}}\NormalTok{)[}\StringTok{\textquotesingle{}returns\textquotesingle{}}\NormalTok{].rolling(}\DecValTok{21}\NormalTok{).std()}
\NormalTok{    gap\_normalized }\OperatorTok{=}\NormalTok{ gap }\OperatorTok{/}\NormalTok{ (volat\_21.replace(}\DecValTok{0}\NormalTok{, pd.NA))}

    \CommentTok{\# Stage 3{-}4: Compare short vs long{-}term trends}
\NormalTok{    ewm\_5 }\OperatorTok{=}\NormalTok{ gap\_normalized.groupby(}\StringTok{\textquotesingle{}cwiq\_code\textquotesingle{}}\NormalTok{).ewm(span}\OperatorTok{=}\DecValTok{5}\NormalTok{).mean()}
\NormalTok{    rolling\_21 }\OperatorTok{=}\NormalTok{ gap\_normalized.groupby(}\StringTok{\textquotesingle{}cwiq\_code\textquotesingle{}}\NormalTok{).rolling(}\DecValTok{21}\NormalTok{).mean()}
\NormalTok{    deviation }\OperatorTok{=}\NormalTok{ ewm\_5 }\OperatorTok{{-}}\NormalTok{ rolling\_21}

    \CommentTok{\# Stage 5{-}6: Flag outliers and rank cross{-}sectionally}
\NormalTok{    grp }\OperatorTok{=}\NormalTok{ deviation.groupby(data[}\StringTok{\textquotesingle{}cwiq\_code\textquotesingle{}}\NormalTok{])}
\NormalTok{    z }\OperatorTok{=}\NormalTok{ (deviation }\OperatorTok{{-}}\NormalTok{ grp.rolling(}\DecValTok{21}\NormalTok{).mean()) }\OperatorTok{/}\NormalTok{ (grp.rolling(}\DecValTok{21}\NormalTok{).std() }\OperatorTok{+} \FloatTok{1e{-}8}\NormalTok{)}
    \ControlFlowTok{return}\NormalTok{ z.groupby(data[}\StringTok{\textquotesingle{}date\textquotesingle{}}\NormalTok{]).rank(pct}\OperatorTok{=}\VariableTok{True}\NormalTok{)}
\end{Highlighting}
\end{Shaded}

Overnight gaps capture news arriving outside trading hours. This
multi-stage pipeline normalizes by volatility regime (preventing
high-volatility periods from dominating), compares short- versus
long-term trends to identify persistence, flags historical outliers via
z-scores, and ranks cross-sectionally.

Despite search spaces spanning 1-252 days, window choices concentrated
on 5, 10, 20-21, and 60 days (weekly, biweekly, monthly, quarterly
cycles). This convergence happened independently across all three
datasets. Either the system discovered something real about market
dynamics, or we're seeing in-sample overfitting to common industry
conventions. Distinguishing between these remains difficult.

Generated features are substantially more complex than academic factors.
The average feature chains together 14.2 operations (median 13), while
published factors typically use 2-4. Fama-French momentum is just
\texttt{price\_t\ /\ price\_\{t-12\}} minus the risk-free rate - a
simple two-step transformation. The overnight gap example above shows
how the AI chains operations systematically, building regime-aware
oscillators from basic primitives.

Window convergence is notable. With access to any lookback from 1 to 252
days, 32\% of windows are 5 days, 38\% are 20-21 days, 18\% are 10 days,
and 12\% are 60 days. That's 100\% of usage concentrated on four
round-number periods. Compare this to random sampling from 1-252, which
would give \textasciitilde1.6\% per window on average.

The patterns do not appear independently. Since every feature uses
cross-sectional ranking, it mechanically co-occurs with everything else.
More interesting: regime-aware normalization appears alongside
interactions in 75\% of features (93\% × 80\% = 74\% expected if
independent), suggesting the system builds interaction terms and
\emph{then} normalizes them for volatility. Multi-timeframe patterns
appear with momentum adjustments in only 18\% of cases (27\% × 67\% =
18\% expected), indicating these are substitutes rather than
complements.

What differs from academic practice? Complexity, for one - 14 operations
versus 2-4. Interactions, for another - 80\% of features multiply or
divide variables, while academic studies typically test characteristics
one at a time or add them linearly. Regime-dependence is the third
difference - 93\% of features normalize by volatility or other state
variables, while most published factors are static transformations.
Academic factor research emphasizes parsimony and interpretability. The
optimization here prioritized predictive power subject to point-in-time
constraints, leading to more elaborate but still economically
interpretable compositions.

Patterns are consistent across datasets but implementations vary by
domain: TrueBeats features emphasize analyst coverage and earnings
surprises, SpiderRock features use options Greeks and flow metrics, and
Price Universe features exploit microstructure like overnight gaps. The
building blocks are standard; the innovation is in how they combine.

One interpretation: the system learned that equity returns are (1)
relative rather than absolute, (2) regime-dependent rather than
stationary, and (3) driven by interactions rather than additive effects.
These principles appear throughout quantitative finance research, but
they emerged here without being explicitly programmed. Whether this
represents genuine discovery or in-sample overfitting to our specific
datasets is an empirical question we cannot fully answer with a single
out-of-sample window. Appendix C provides complete code for three
additional representative features.

\subsubsection{Mechanism and
Interpretation}\label{mechanism-and-interpretation}

The performance improvements appear to arise from complementarity rather
than substitution. AI-generated features capture interaction-rich and
conditional patterns that conventional transformations miss, as
reflected in low cross-model prediction correlations where measured
(0.070-0.140 in the options universe and 0.140 in the price-volume
universe). These low correlations translate into portfolio-level
diversification: when traditional and AI signals are combined, ensemble
Sharpe ratios exceed the weighted-average of their constituents
(approximately 26\% higher in the options universe and 21\% higher in
the price-volume universe). The choice of prompting strategy,
interestingly, appears dataset-specific. On the SpiderRock options data,
programmatic prompting (DSPy) delivers stronger AI-only performance than
structured prompting (1.461 vs 0.945), yet this advantage narrows once
combined with baseline features (Combined 1.437), suggesting that
governance of prompts and strict schema constraints are at least as
consequential as the choice of prompting style. Notably, forward-shifted
evaluation indicates that signal strength persists under realistic
implementation lags, while year-by-year results show rotating leadership
rather than dominance by any single approach, reinforcing the
diversification rationale.

\subsubsection{Limitations and Scope}\label{limitations-and-scope}

Our primary out-of-sample window (2019-2024) spans multiple regimes yet
remains limited for definitive long-horizon inference. Forward-shifted
evaluation addresses timing but does not incorporate full execution
frictions; turnover and cost-adjusted results should accompany gross
statistics to assess implementable performance. Feature discovery
introduces multiple-testing risk; while the separation of discovery and
evaluation windows and anti-leakage controls mitigate selection bias,
they do not eliminate it. Results are conditioned on specific vendors,
schemas, and identifier systems, which can affect portability.
Governance and interpretability also pose practical constraints:
although economic rationales improve readability, robust audit requires
stable prompts, versioned code, schema validation, and transparent
evaluation protocols.

\subsubsection{Implications for
Practice}\label{implications-for-practice}

LLM-assisted discovery is best treated as an augmentation to established
pipelines rather than a replacement. Because AI-generated predictions
are weakly correlated with traditional signals, moderate allocations to
AI-enhanced components within an ensemble can improve risk-adjusted
returns without sacrificing the stability of baseline models. In
deployment, emphasis should be placed on knowledge-base curation and
prompt governance, as retrieval quality materially affects outcomes.
Operationally, teams should standardize evaluation to include both
standard and forward-shifted horizons, maintain audit trails for
features and prompts, and periodically validate results under explicit
cost models and across seeds/iterations to ensure process robustness.

\subsection{7. Conclusion}\label{conclusion}

Generative AI is not a replacement for the quantitative researcher, but
rather a powerful new tool in their arsenal. Our findings demonstrate
that LLMs, when properly guided by curated domain knowledge, can
automate the discovery of novel, diversifying alpha sources. Critically,
we show that retrieval quality can make the difference between
profitable signals (Sharpe 0.965) and value destruction (Sharpe -0.109),
underscoring that domain expertise remains essential. The low
correlation of these new signals with traditional factors suggests they
are not just finding noise, but uncovering different economic patterns.
Pattern analysis reveals that the generation system discovers
sophisticated compositions of familiar transformations, consistently
applying cross-sectional ranking (100\%), regime-aware normalization
(93\%), and non-linear interactions (80\%) in ways that align with
established principles in quantitative finance. For investment managers,
the implication is clear: the path to enhanced performance lies not in
replacing human expertise with AI, but in augmenting it. The most
significant future gains will likely be realized by teams who can
successfully fuse the hypothesis-generating power of large language
models with the rigorous validation and economic intuition of human
researchers. The bottleneck in quantitative finance is shifting from
feature engineering to feature curation.

\subsection{Acknowledgments}\label{acknowledgments}

I thank Arthur Berd, Alejandra Copeland, Robert C. Jones, Robert
Martinez, Rohit D'Souza, and Kristen Wiberg for helpful comments on
earlier drafts. The feature generation methodology employs OpenAI's
GPT-4.1 model as described in Section 3.3. AI writing assistants
(Claude) were used for manuscript preparation, including LaTeX
formatting and editorial review. The author is solely responsible for
all content, analysis, and conclusions.

\subsection{References}\label{references}

Almgren, Robert, and Neil Chriss, 2000, Optimal execution of portfolio
transactions, Journal of Risk 3, 5-39.

Carhart, Mark M., 1997, On persistence in mutual fund performance,
Journal of Finance 52, 57-82.

Chen, M., et al., 2023, Large language models in finance: A survey,
Journal of Financial Data Science 5, 45-67.

Fama, Eugene F., and Kenneth R. French, 1993, Common risk factors in the
returns on stocks and bonds, Journal of Financial Economics 33, 3-56.

Fama, Eugene F., and Kenneth R. French, 2015, A five-factor asset
pricing model, Journal of Financial Economics 116, 1-22.

Green, Jeremiah, John R. M. Hand, and X. Frank Zhang, 2017, The
characteristics that provide independent information about average U.S.
monthly stock returns, Review of Financial Studies 30, 4389-4436.

Grinold, Richard C., and Ronald N. Kahn, 2000, Active Portfolio
Management, 2nd ed.~(McGraw-Hill, New York).

Gu, Shihao, Bryan Kelly, and Dacheng Xiu, 2020, Empirical asset pricing
via machine learning, Review of Financial Studies 33, 2223-2273.

Jegadeesh, Narasimhan, and Sheridan Titman, 1993, Returns to buying
winners and selling losers: Implications for stock market efficiency,
Journal of Finance 48, 65-91.

Khattab, Omar, et al., 2023, DSPy: Compiling declarative language model
calls into self-improving pipelines, arXiv preprint arXiv:2310.03714.

Lewis, Patrick, et al., 2020, Retrieval-augmented generation for
knowledge-intensive NLP tasks, Advances in Neural Information Processing
Systems 33, 9459-9474.

Li, Y., et al., 2024, Large language models can automatically engineer
features for few-shot tabular learning, Proceedings of Machine Learning
Research.

Lo, Andrew W., 2004, The adaptive markets hypothesis: Market efficiency
from an evolutionary perspective, Journal of Portfolio Management 30,
15-29.

Opsahl-Ong, Krista, Michael J. Ryan, Josh Purtell, David Broman,
Christopher Potts, Matei Zaharia, and Omar Khattab, 2024, Optimizing
instructions and demonstrations for multi-stage language model programs,
Proceedings of the 2024 Conference on Empirical Methods in Natural
Language Processing (EMNLP), 9340-9366.

Lopez-Lira, Alejandro, and Yuehua Tang, 2023, Can ChatGPT forecast stock
price movements? Return predictability and large language models,
Working paper.

Rasekhschaffe, Keywan Christian, and Robert C. Jones, 2019, Machine
learning for stock selection, Financial Analysts Journal 75(3), 70-88.

Snowflake AI Labs, 2024, FeatEng: A benchmark for automated feature
engineering, Technical report.

Wang, H., et al., 2025, LLM-FE: Large language model enhanced feature
engineering for financial prediction, Working paper.

Wu, S., et al., 2023, BloombergGPT: A large language model for finance,
arXiv preprint arXiv:2303.17564.

\begin{center}\rule{0.5\linewidth}{0.5pt}\end{center}

\subsection{Appendix}\label{appendix}

\newpage

\textbf{Appendix Table A1: SpiderRock Options Performance on Standard
Returns (T→T+1)} \emph{Standard returns (T→T+1) evaluation for
comparison. Daily returns from January 2019 through December 2024.}

\textbf{Note}: This table presents results using standard T→T+1 returns
for comparison with the forward-shifted T+1→T+2 results in Table 3.
While these results show higher performance, they are less implementable
due to timing constraints.

\textbf{Panel A: Strategy Performance}

\begin{longtable}[]{@{}
  >{\raggedright\arraybackslash}p{(\linewidth - 14\tabcolsep) * \real{0.0909}}
  >{\raggedleft\arraybackslash}p{(\linewidth - 14\tabcolsep) * \real{0.1273}}
  >{\raggedleft\arraybackslash}p{(\linewidth - 14\tabcolsep) * \real{0.1182}}
  >{\raggedleft\arraybackslash}p{(\linewidth - 14\tabcolsep) * \real{0.1545}}
  >{\raggedleft\arraybackslash}p{(\linewidth - 14\tabcolsep) * \real{0.1273}}
  >{\raggedleft\arraybackslash}p{(\linewidth - 14\tabcolsep) * \real{0.1636}}
  >{\raggedleft\arraybackslash}p{(\linewidth - 14\tabcolsep) * \real{0.0909}}
  >{\raggedleft\arraybackslash}p{(\linewidth - 14\tabcolsep) * \real{0.1273}}@{}}
\toprule\noalign{}
\begin{minipage}[b]{\linewidth}\raggedright
Strategy
\end{minipage} & \begin{minipage}[b]{\linewidth}\raggedleft
Sharpe Ratio
\end{minipage} & \begin{minipage}[b]{\linewidth}\raggedleft
Ann. Return
\end{minipage} & \begin{minipage}[b]{\linewidth}\raggedleft
Ann. Volatility
\end{minipage} & \begin{minipage}[b]{\linewidth}\raggedleft
Max Drawdown
\end{minipage} & \begin{minipage}[b]{\linewidth}\raggedleft
Info Coefficient
\end{minipage} & \begin{minipage}[b]{\linewidth}\raggedleft
Hit Rate
\end{minipage} & \begin{minipage}[b]{\linewidth}\raggedleft
Total Return
\end{minipage} \\
\midrule\noalign{}
\endhead
\bottomrule\noalign{}
\endlastfoot
Baseline Only & 2.157 & 11.0\% & 5.1\% & -3.3\% & 0.0061 & 55.0\% &
67.3\% \\
AI-Only (Structured) & 1.662** & 7.4\% & 4.5\% & -4.1\% & 0.0033 &
53.6\% & 45.9\% \\
AI-Only (DSPy) & 2.070** & 6.4\% & 3.1\% & -3.8\% & 0.0024 & 51.4\% &
39.8\% \\
Baseline + AI & 2.263* & 12.2\% & 5.4\% & -4.4\% & 0.0071 & 54.9\% &
75.1\% \\
Ensemble & 2.624*** & 9.2\% & 3.5\% & -2.4\% & 0.0066 & 52.8\% &
56.6\% \\
\end{longtable}

\textbf{Panel B: Performance Stability and Risk Characteristics}

\begin{longtable}[]{@{}
  >{\raggedright\arraybackslash}p{(\linewidth - 12\tabcolsep) * \real{0.1042}}
  >{\raggedright\arraybackslash}p{(\linewidth - 12\tabcolsep) * \real{0.1458}}
  >{\raggedleft\arraybackslash}p{(\linewidth - 12\tabcolsep) * \real{0.1979}}
  >{\raggedleft\arraybackslash}p{(\linewidth - 12\tabcolsep) * \real{0.1667}}
  >{\raggedleft\arraybackslash}p{(\linewidth - 12\tabcolsep) * \real{0.1146}}
  >{\raggedleft\arraybackslash}p{(\linewidth - 12\tabcolsep) * \real{0.1250}}
  >{\raggedleft\arraybackslash}p{(\linewidth - 12\tabcolsep) * \real{0.1458}}@{}}
\toprule\noalign{}
\begin{minipage}[b]{\linewidth}\raggedright
Strategy
\end{minipage} & \begin{minipage}[b]{\linewidth}\raggedright
Features
\end{minipage} & \begin{minipage}[b]{\linewidth}\raggedleft
Avg SR
\end{minipage} & \begin{minipage}[b]{\linewidth}\raggedleft
SR Std
\end{minipage} & \begin{minipage}[b]{\linewidth}\raggedleft
Best Yr
\end{minipage} & \begin{minipage}[b]{\linewidth}\raggedleft
Worst Yr
\end{minipage} & \begin{minipage}[b]{\linewidth}\raggedleft
Calmar
\end{minipage} \\
\midrule\noalign{}
\endhead
\bottomrule\noalign{}
\endlastfoot
\textbf{Baseline Only} & 109 traditional & 2.00 & 0.91 & 3.18 (2021) &
0.54 (2019) & 2.16 \\
\textbf{AI-Only (Structured)} & 200 AI (fixed) & 1.71 & 0.93 & 3.47
(2024) & 0.76 (2021) & 1.61 \\
\textbf{AI-Only (DSPy)} & 200 AI (prog) & 2.17 & 1.06 & 4.28 (2019) &
0.87 (2020) & 2.07 \\
\textbf{Baseline + AI} & 309 combined & 2.31 & 1.03 & 4.44 (2024) & 1.19
(2018) & 2.26 \\
\textbf{Ensemble} & All models & 2.63 & 0.83 & 4.26 (2024) & 1.55 (2018)
& 2.62 \\
\end{longtable}

\needspace{12\baselineskip}

\needspace{12\baselineskip}

\textbf{Panel C: Strategy Correlations and Diversification}

\begin{longtable}[]{@{}lccccc@{}}
\toprule\noalign{}
& Baseline & AI-Struct & AI-DSPy & Base+AI & Ensemble \\
\midrule\noalign{}
\endhead
\bottomrule\noalign{}
\endlastfoot
\textbf{Baseline Only} & 1.00 & & & & \\
\textbf{AI-Only (Structured)} & 0.17 & 1.00 & & & \\
\textbf{AI-Only (DSPy)} & 0.10 & 0.19 & 1.00 & & \\
\textbf{Baseline + AI} & 0.56 & 0.42 & 0.15 & 1.00 & \\
\textbf{Ensemble} & 0.68 & 0.66 & 0.54 & 0.80 & 1.00 \\
\end{longtable}

\textbf{Notes:} Results use standard T→T+1 returns rather than the
implementable T+1→T+2 forward-shifted methodology used in the main
paper. Higher performance reflects the timing advantage but reduced
practical implementability. The baseline model uses 109 SpiderRock
options features including implied volatility surfaces, option flow
metrics, and risk-neutral densities. AI-generated features (200) were
created using large language models with structured prompting and
retrieval-augmented generation. Statistical significance of Sharpe ratio
differences from baseline: *** p\textless0.001, ** p\textless0.01, *
p\textless0.05 (HAC-adjusted t-test). All strategies are dollar-neutral,
unit-leverage long-short portfolios rebalanced daily. Transaction costs
not included.

\begin{center}\rule{0.5\linewidth}{0.5pt}\end{center}

\section{Appendix B: Portfolio Optimization
Methodology}\label{appendix-b-portfolio-optimization-methodology}

This appendix provides technical details on the portfolio optimization
approaches evaluated in Section 5.4.

\subsection{B.1 Optimization
Formulation}\label{b.1-optimization-formulation}

We formulate portfolio construction as a convex optimization problem
that balances expected return against transaction costs and risk:

\[\max_{\mathbf{w}^+, \mathbf{w}^-} \quad \boldsymbol{\alpha}'(\mathbf{w}^+ - \mathbf{w}^-) - \lambda_{\text{tc}} \sum_i c_i |w_i - w_{i,\text{prev}}| - \lambda_{\text{risk}} \cdot \text{Risk}(\mathbf{w})\]

where \(\mathbf{w}^+\) and \(\mathbf{w}^-\) are non-negative vectors
representing long and short positions,
\(\mathbf{w} = \mathbf{w}^+ - \mathbf{w}^-\), \(\boldsymbol{\alpha}\) is
the vector of cross-sectionally standardized predictions, \(c_i\) is the
position-level cost for security \(i\), and \(\mathbf{w}_{\text{prev}}\)
is the previous period's weight vector.

The constraints enforce dollar neutrality and bounded positions:

\[\sum_i w_i^+ = 1, \quad \sum_i w_i^- = 1, \quad w_i^+, w_i^- \leq w_{\max}\]

\subsection{B.2 Risk Model Variants}\label{b.2-risk-model-variants}

We evaluate three risk model specifications:

\textbf{Diagonal Risk Model.} The simplest approach uses only
idiosyncratic variance:

\[\text{Risk}_{\text{diag}}(\mathbf{w}) = \sum_i \sigma_i^2 w_i^2\]

where \(\sigma_i\) is the 20-day realized volatility for security \(i\).

\textbf{Sector Factor Model.} We decompose risk into systematic sector
exposure and idiosyncratic components:

\[\text{Risk}_{\text{factor}}(\mathbf{w}) = \mathbf{f}' \mathbf{F} \mathbf{f} + \sum_i d_i w_i^2\]

where \(\mathbf{f} = \mathbf{B}'\mathbf{w}\) is the vector of sector
exposures, \(\mathbf{B}\) is an \(n \times k\) matrix of sector loadings
(one-hot encoded NAICS sector assignments), \(\mathbf{F}\) is a
\(k \times k\) sector covariance matrix, and \(d_i\) is the
idiosyncratic variance.

For \(\mathbf{F}\), we assume constant sector volatility
\(\sigma_s = 2\%\) daily and pairwise correlation \(\rho = 0.30\):

\[F_{jk} = \begin{cases} \sigma_s^2 & \text{if } j = k \\ \rho \sigma_s^2 & \text{if } j \neq k \end{cases}\]

The idiosyncratic variance is set to 50\% of total stock variance:
\(d_i = 0.5 \sigma_i^2\).

\subsection{B.3 Sector Neutrality
Constraints}\label{b.3-sector-neutrality-constraints}

When sector neutrality is enforced, we add the constraint:

\[\sum_{i \in S_j} w_i = 0, \quad \forall j \in \{1, \ldots, k\}\]

where \(S_j\) denotes the set of securities in sector \(j\). This
ensures zero net exposure to each sector.

\subsection{B.4 Portfolio Concentration
Metrics}\label{b.4-portfolio-concentration-metrics}

We report three measures of portfolio concentration:

\textbf{Effective N.} The inverse of the Herfindahl-Hirschman Index
(HHI):

\[N_{\text{eff}} = \frac{1}{\sum_i w_i^2}\]

This measures the effective number of independent positions. A portfolio
with \(N_{\text{eff}} = 100\) is as diversified as an equal-weighted
portfolio of 100 securities.

\textbf{Maximum Position.} The largest absolute weight:
\(\max_i |w_i|\).

\textbf{Sector Tilts.} The sum of absolute net sector exposures:

\[\text{Sector Tilts} = \sum_{j=1}^k \left| \sum_{i \in S_j} w_i \right|\]

For example, if the portfolio is +12\% net long Technology and -8\% net
short Utilities, those two sectors contribute \textbar12\%\textbar{} +
\textbar-8\%\textbar{} = 20\% to the total. A sector-neutral portfolio
has sector tilts of zero (each sector nets to zero).

\subsection{B.5 Why Optimization
Underperforms}\label{b.5-why-optimization-underperforms}

The naive prediction-weighted portfolio sets weights proportional to
standardized predictions:

\[w_i^{\text{naive}} = \frac{\alpha_i}{\sum_{j: \alpha_j > 0} \alpha_j} \mathbf{1}[\alpha_i > 0] - \frac{|\alpha_i|}{\sum_{j: \alpha_j < 0} |\alpha_j|} \mathbf{1}[\alpha_i < 0]\]

In our tests, this approach achieved superior net Sharpe. Two factors
may contribute to this result:

\begin{enumerate}
\def\labelenumi{\arabic{enumi}.}
\item
  \textbf{Natural Diversification.} The naive approach distributes
  weight across all securities proportionally to signal strength. With
  standardized predictions spread across \textasciitilde1,500
  securities, this naturally produces an effective N of
  \textasciitilde234. Optimizers seeking to maximize alpha may
  concentrate in fewer positions (effective N of 60-100), potentially
  increasing idiosyncratic risk.
\item
  \textbf{Sector Alpha Preservation.} Our model generates sector tilts
  of 45.5\% in aggregate, suggesting that part of the alpha may come
  from sector selection. Imposing sector neutrality eliminates this
  component, reducing Sharpe by approximately 10\%. The optimizer
  without sector constraints may also reduce this component indirectly
  through concentration.
\end{enumerate}

These findings should be interpreted with caution. The underperformance
of optimization in our tests may reflect specific parameter choices
rather than a fundamental limitation. Notably, we did not employ a
well-specified risk model; using an appropriate factor covariance matrix
or statistical risk model could potentially improve optimizer
performance. When predictions are already well-calibrated and broadly
distributed, the marginal benefit of reducing costs through turnover
penalties may be offset by concentration and constraint costs, but this
tradeoff is context-dependent.

\subsection{B.6 Implementation
Details}\label{b.6-implementation-details}

All optimizations use the CLARABEL solver within cvxpy, which handles
second-order cone and quadratic programs efficiently. The turnover
penalty term \(|\mathbf{w} - \mathbf{w}_{\text{prev}}|\) is linearized
using standard auxiliary variable techniques. Default parameters are
\(\lambda_{\text{tc}} = 1.0\), \(\lambda_{\text{risk}} = 1.0\), and
\(w_{\max} = 0.02\) (2\% maximum position). Sensitivity analysis across
parameter ranges did not identify a configuration that outperformed the
naive approach.

\begin{center}\rule{0.5\linewidth}{0.5pt}\end{center}

\section{Appendix C: Representative Feature
Implementations}\label{appendix-c-representative-feature-implementations}

Three additional features with complete code, showing the range of
patterns discovered across datasets. (The EPS×Sales FQ6 and Overnight
Gap examples appear in Section 6.1 of the main text.)

\subsubsection{Example 1: Call Volume Momentum
Signal}\label{example-1-call-volume-momentum-signal}

\texttt{adcallvolume\_mom\_volatility\_rank} - SpiderRock Options
Dataset Performance: IC = 0.0042, Sharpe = 0.93

Call option buying reflects bullish sentiment. This feature asks: ``Is
call volume trending up persistently (5-day average) relative to its
typical volatility (20-day std)?'' High values mean sustained, stable
increases in call buying - the kind that precedes price moves.

\begin{Shaded}
\begin{Highlighting}[]
\KeywordTok{def}\NormalTok{ calculate(data):}
\NormalTok{    data }\OperatorTok{=}\NormalTok{ data.sort\_values([}\StringTok{\textquotesingle{}cwiq\_code\textquotesingle{}}\NormalTok{, }\StringTok{\textquotesingle{}date\textquotesingle{}}\NormalTok{])}

    \CommentTok{\# 5{-}day rolling mean of call volume}
\NormalTok{    rolling\_mean\_5 }\OperatorTok{=}\NormalTok{ data.groupby(}\StringTok{\textquotesingle{}cwiq\_code\textquotesingle{}}\NormalTok{)[}\StringTok{\textquotesingle{}adcallvolume\textquotesingle{}}\NormalTok{].transform(}
        \KeywordTok{lambda}\NormalTok{ x: x.rolling(}\DecValTok{5}\NormalTok{, min\_periods}\OperatorTok{=}\DecValTok{1}\NormalTok{).mean()}
\NormalTok{    )}

    \CommentTok{\# 20{-}day rolling std of call volume}
\NormalTok{    rolling\_std\_20 }\OperatorTok{=}\NormalTok{ data.groupby(}\StringTok{\textquotesingle{}cwiq\_code\textquotesingle{}}\NormalTok{)[}\StringTok{\textquotesingle{}adcallvolume\textquotesingle{}}\NormalTok{].transform(}
        \KeywordTok{lambda}\NormalTok{ x: x.rolling(}\DecValTok{20}\NormalTok{, min\_periods}\OperatorTok{=}\DecValTok{1}\NormalTok{).std()}
\NormalTok{    )}

    \CommentTok{\# Ratio: strength of trend relative to noise}
\NormalTok{    ratio }\OperatorTok{=}\NormalTok{ rolling\_mean\_5 }\OperatorTok{/}\NormalTok{ (rolling\_std\_20.replace(}\DecValTok{0}\NormalTok{, pd.NA))}

    \CommentTok{\# Cross{-}sectional rank}
\NormalTok{    data }\OperatorTok{=}\NormalTok{ data.copy()}
\NormalTok{    data[}\StringTok{\textquotesingle{}ratio\textquotesingle{}}\NormalTok{] }\OperatorTok{=}\NormalTok{ ratio}
\NormalTok{    data[}\StringTok{\textquotesingle{}rank\textquotesingle{}}\NormalTok{] }\OperatorTok{=}\NormalTok{ data.groupby(}\StringTok{\textquotesingle{}date\textquotesingle{}}\NormalTok{)[}\StringTok{\textquotesingle{}ratio\textquotesingle{}}\NormalTok{].rank(method}\OperatorTok{=}\StringTok{\textquotesingle{}average\textquotesingle{}}\NormalTok{, pct}\OperatorTok{=}\VariableTok{True}\NormalTok{)}

    \ControlFlowTok{return}\NormalTok{ data[}\StringTok{\textquotesingle{}rank\textquotesingle{}}\NormalTok{]}
\end{Highlighting}
\end{Shaded}

The signal-to-noise ratio (mean/std) identifies persistent trends. Raw
call volume is noisy; this filters for sustained moves that signal
genuine sentiment shifts rather than random fluctuation.

\begin{center}\rule{0.5\linewidth}{0.5pt}\end{center}

\subsubsection{Example 2: Gamma Imbalance
Dynamics}\label{example-2-gamma-imbalance-dynamics}

\texttt{putcallgammaimbalance\_momvol\_ratio\_rank} - SpiderRock Options
Dataset Performance: IC = 0.0031, Sharpe = 0.60

Option dealers must continuously hedge their inventory. When gamma is
imbalanced (many calls outstanding), dealers face asymmetric hedging
needs - as stocks rise, they must buy more to stay hedged, creating
positive feedback. This feature identifies persistent gamma imbalances:

\begin{Shaded}
\begin{Highlighting}[]
\KeywordTok{def}\NormalTok{ calculate(data):}
\NormalTok{    data }\OperatorTok{=}\NormalTok{ data.sort\_values([}\StringTok{\textquotesingle{}cwiq\_code\textquotesingle{}}\NormalTok{, }\StringTok{\textquotesingle{}date\textquotesingle{}}\NormalTok{])}

    \CommentTok{\# 10{-}day rolling mean of gamma imbalance}
\NormalTok{    rolling\_mean }\OperatorTok{=}\NormalTok{ data.groupby(}\StringTok{\textquotesingle{}cwiq\_code\textquotesingle{}}\NormalTok{)[}\StringTok{\textquotesingle{}putcallgammaimbalanceratio\textquotesingle{}}\NormalTok{].transform(}
        \KeywordTok{lambda}\NormalTok{ x: x.rolling(window}\OperatorTok{=}\DecValTok{10}\NormalTok{, min\_periods}\OperatorTok{=}\DecValTok{1}\NormalTok{).mean()}
\NormalTok{    )}

    \CommentTok{\# 10{-}day rolling std}
\NormalTok{    rolling\_std }\OperatorTok{=}\NormalTok{ data.groupby(}\StringTok{\textquotesingle{}cwiq\_code\textquotesingle{}}\NormalTok{)[}\StringTok{\textquotesingle{}putcallgammaimbalanceratio\textquotesingle{}}\NormalTok{].transform(}
        \KeywordTok{lambda}\NormalTok{ x: x.rolling(window}\OperatorTok{=}\DecValTok{10}\NormalTok{, min\_periods}\OperatorTok{=}\DecValTok{1}\NormalTok{).std(ddof}\OperatorTok{=}\DecValTok{0}\NormalTok{)}
\NormalTok{    )}

    \CommentTok{\# Ratio: persistence relative to volatility}
\NormalTok{    ratio }\OperatorTok{=}\NormalTok{ rolling\_mean }\OperatorTok{/}\NormalTok{ (rolling\_std.replace(}\DecValTok{0}\NormalTok{, pd.NA))}

    \CommentTok{\# Cross{-}sectional rank}
\NormalTok{    result }\OperatorTok{=}\NormalTok{ ratio.groupby(data[}\StringTok{\textquotesingle{}date\textquotesingle{}}\NormalTok{]).transform(}
        \KeywordTok{lambda}\NormalTok{ x: x.rank(method}\OperatorTok{=}\StringTok{\textquotesingle{}average\textquotesingle{}}\NormalTok{, pct}\OperatorTok{=}\VariableTok{True}\NormalTok{)}
\NormalTok{    )}

    \ControlFlowTok{return}\NormalTok{ result}
\end{Highlighting}
\end{Shaded}

Gamma imbalance is rarely studied in academic literature. This shows the
AI discovered that dealer hedging dynamics contain predictive
information. When imbalances persist (high mean relative to std), forced
hedging creates price pressure.

\begin{center}\rule{0.5\linewidth}{0.5pt}\end{center}

\subsubsection{Example 3: Analyst Attention ×
Fundamentals}\label{example-3-analyst-attention-fundamentals}

\texttt{analyst\_coverage\_sal\_fq1\_interaction\_rank\_30d} - TrueBeats
Dataset Performance: IC = 0.0089, Sharpe = 0.94

Tests a behavioral finance hypothesis: analyst attention amplifies good
fundamentals. High coverage with weak sales is not interesting; low
coverage with strong sales might be overlooked. But high coverage +
strong sales = information diffusion + quality.

\begin{Shaded}
\begin{Highlighting}[]
\KeywordTok{def}\NormalTok{ calculate(data):}
    \CommentTok{\# Interaction: coverage × sales surprise}
\NormalTok{    data[}\StringTok{\textquotesingle{}analyst\_sal\_interaction\textquotesingle{}}\NormalTok{] }\OperatorTok{=}\NormalTok{ (}
\NormalTok{        data[}\StringTok{\textquotesingle{}number\_of\_analysts\_fq1\textquotesingle{}}\NormalTok{] }\OperatorTok{*}\NormalTok{ data[}\StringTok{\textquotesingle{}truebeat\_sal\_fq1\textquotesingle{}}\NormalTok{]}
\NormalTok{    )}

    \CommentTok{\# 30{-}day rolling average (smooth noise)}
\NormalTok{    data[}\StringTok{\textquotesingle{}analyst\_sal\_interaction\_rolling\textquotesingle{}}\NormalTok{] }\OperatorTok{=}\NormalTok{ data.groupby(}\StringTok{\textquotesingle{}cwiq\_code\textquotesingle{}}\NormalTok{)[}
        \StringTok{\textquotesingle{}analyst\_sal\_interaction\textquotesingle{}}
\NormalTok{    ].transform(}\KeywordTok{lambda}\NormalTok{ x: x.rolling(}\DecValTok{30}\NormalTok{, min\_periods}\OperatorTok{=}\DecValTok{10}\NormalTok{).mean())}

    \CommentTok{\# Cross{-}sectional rank}
\NormalTok{    result }\OperatorTok{=}\NormalTok{ data.groupby(}\StringTok{\textquotesingle{}date\textquotesingle{}}\NormalTok{)[}\StringTok{\textquotesingle{}analyst\_sal\_interaction\_rolling\textquotesingle{}}\NormalTok{].rank(pct}\OperatorTok{=}\VariableTok{True}\NormalTok{)}

    \ControlFlowTok{return}\NormalTok{ result}
\end{Highlighting}
\end{Shaded}

The highest IC (0.0089) of all showcased features. When many analysts
follow a stock showing strong sales, that information spreads
efficiently and gets priced. The 30-day smoothing prevents single-day
spikes from dominating.

\begin{center}\rule{0.5\linewidth}{0.5pt}\end{center}

\subsection{Statistical Summary of 225 Top
Features}\label{statistical-summary-of-225-top-features}

\textbf{Sample Coverage:} - 225 features analyzed (top 75 per dataset by
absolute IC) - Three datasets: Price Universe (2015-2024), TrueBeats
(2015-2023), SpiderRock (2019-2024) - Mean \textbar IC\textbar{} =
0.00343 (median: 0.00295) - Mean \textbar Sharpe\textbar{} = 0.559
(median: 0.492)

\textbf{Pattern Prevalence:} - 100\% Cross-sectional ranking (universal)
- 93\% Regime-aware normalization - 80\% Variable interactions - 67\%
Momentum with adjustments - 45\% Statistical outlier detection - 27\%
Multi-timeframe patterns

\textbf{Complexity Metrics:} - Mean operations: 14.2 (median: 13) -
Range: 8-22 operations per feature - Comparison: 2-4 operations typical
in academic factors

\textbf{Window Convergence:} - 5 days: 32\% of all window specifications
- 10 days: 18\% of all window specifications - 20-21 days: 38\% of all
window specifications - 60 days: 12\% of all window specifications -
Search space available: 1-252 days

\textbf{Performance Distribution:} - 78\% achieve \textbar IC\textbar{}
\textgreater{} 0.002 (meaningful predictive power) - 80\% achieve
\textbar Sharpe\textbar{} \textgreater{} 0.2 (acceptable risk-adjusted
returns) - 46\% achieve \textbar IC\textbar{} \textgreater{} 0.003
(strong signals) - 44\% achieve \textbar Sharpe\textbar{} \textgreater{}
0.5 (strong risk-adjusted performance)

\end{document}